\documentclass[a4paper,11pt]{article}
\pdfoutput=1 
\usepackage{jheppub} 
\usepackage[T1]{fontenc} 
\RequirePackage[displaymath]{lineno} 

\usepackage{epsfig}
\usepackage{graphicx}
\usepackage{dcolumn}
\usepackage{bm}
\usepackage{ltablex,booktabs}
\usepackage{overpic}
\usepackage{subfigure}
\usepackage{float}
\usepackage{color}
\usepackage{amsmath}
\usepackage{mathcomp}
\usepackage{mathrsfs}
\usepackage{multirow}
\usepackage{rotating}
\usepackage{amssymb}
\usepackage{gensymb}
\usepackage{amsmath}
\usepackage{tabularx}
\usepackage{epstopdf}
\usepackage{tikz-feynman}
\usepackage{makecell}

\def\MeanBF{\ensuremath{(4.15 \pm 0.07_{\rm{stat.}} \pm 0.17_{\rm{syst.}}) \%}}

\def\BF{\ensuremath{(6.06 \pm 0.04_{\rm{stat.}} \pm 0.07_{\rm{syst.}}) \%}}

\title{Amplitude analysis and branching fraction measurement of the Cabibbo-favored decay \boldmath $D^+ \to K^-\pi^+\pi^+\pi^0$}
\collaborationImg{\includegraphics[width=0.15\textwidth, angle=90]{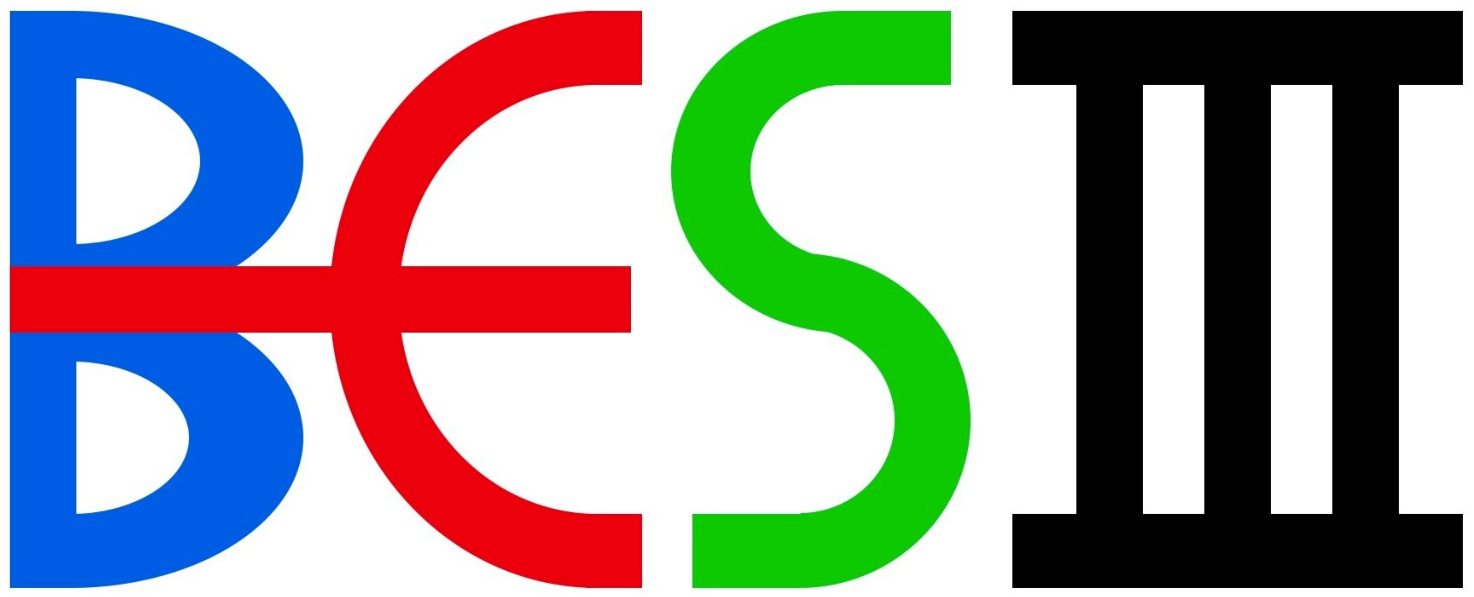}}
\collaboration{BESIII Collaboration}

\date{\today}
\abstract{
An amplitude analysis of the Cabibbo-favored decay $D^+ \to K^-\pi^+\pi^+\pi^0$ is performed, using 7.93 $\rm{fb}^{-1}$ of $e^+e^-$ collision data collected with the BESIII detector at the center-of-mass energy of 3.773 GeV. The branching fractions of the intermediate processes are measured, with the dominant contribution $D^+ \to \bar{K}^{*}(892)^0\rho(770)^+$ observed to have a branching fraction of \MeanBF. With the detection efficiency derived from the amplitude analysis, the absolute branching fraction of $D^+ \to K^-\pi^+\pi^+\pi^0$ is measured to be \BF.}

\arxivnumber{}
\begin{document}
\maketitle
\flushbottom

\section{Introduction}
The Cabibbo-favored~(CF) decay $D^{+}\to K^-\pi^+\pi^+\pi^0$ is  a golden “tag mode” for measurements related to the $D$ meson due to its large branching fraction~(BF) and low background contamination. 
A first measurement of the  BF  and an amplitude analysis of $D^{+}\to K^-\pi^+\pi^+\pi^0$ were performed by the Mark III collaboration~\cite{MK} using a limited data sample. The CLEO collaboration subsequently improved the precision of the BF measurement~\cite{bf}, but did not report any study of the intermediate resonances.

Recently,  the doubly-Cabibbo-suppressed (DCS) decay $D^{+}\to K^+\pi^+\pi^-\pi^0$ was observed for the first time~\cite{PRL125} and its BF was measured to be $(1.13 \pm 0.08_{\rm stat.}\pm0.03_{\rm syst.}) \times 10^{-3}$.  After combining the averaged value of BF for its counterpart CF decay $D^{+}\to K^-\pi^+\pi^+\pi^0$~\cite{PDG}, the ratio of $\frac{\mathcal{B}({D^{+}\to K^+\pi^+\pi^-\pi^0})}{\mathcal{B}({D^{+}\to K^-\pi^+\pi^+\pi^0})}$ is determined to be $(1.81 \pm 0.15)\%$ corresponding to $(6.28 \pm 0.52)\tan^{4}\theta_{C}$, where $\theta_{C}$ is the Cabibbo mixing angle. This ratio is significantly larger than the naive expectation for the DCS rate relative to its CF counterpart decay (0.21-0.58)\%~\cite{PS117,Cheng:2010ry,PRD092003}, which implies that some unknown effects are contributing to  either or both of the two decays and motivates gaining an improved understanding of the resonance structure of $D^{+}\to K^-\pi^+\pi^+\pi^0$. 

\begin{figure}[t]
\centering
    \includegraphics[width=0.45\textwidth]{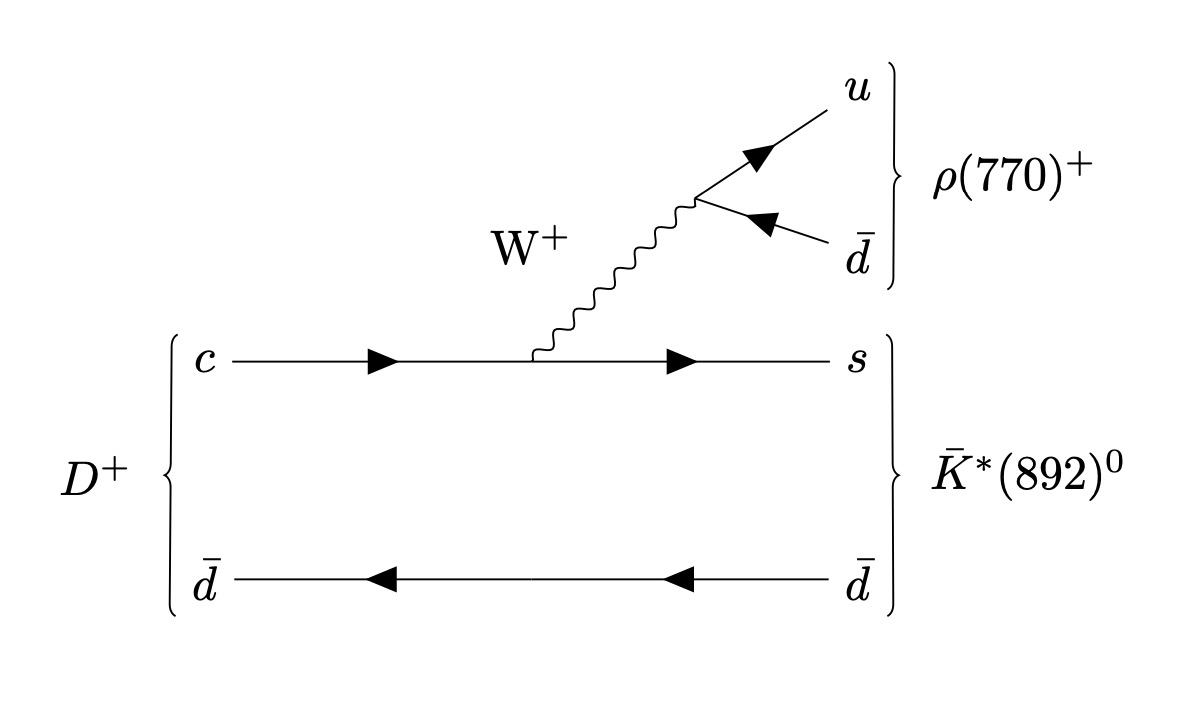}
    \includegraphics[width=0.45\textwidth]{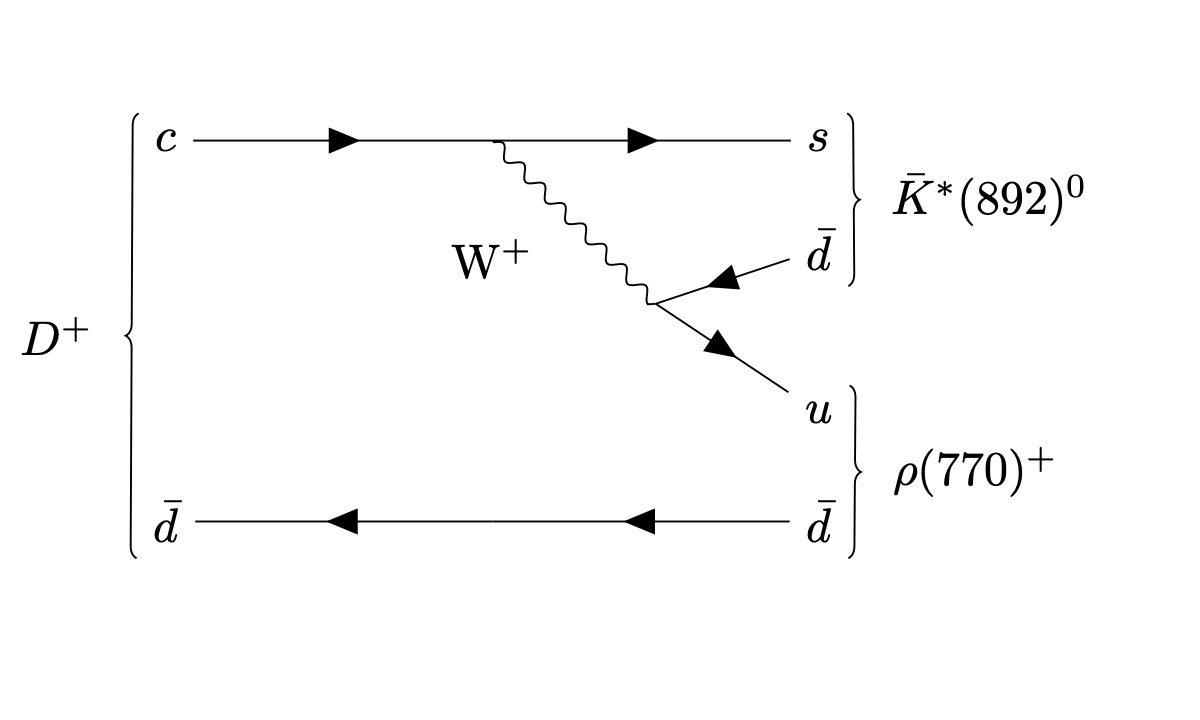}
	        \caption{The Feynman diagrams of $D^+ \to \bar{K}^{*}(892)^0\rho(770)^+$.}
\label{topology1}
\end{figure}

Comparing to the case of many well-studied three-body decays~\cite{PLB653,PRD052001,PRD012006}, an amplitude analysis of $D^{+}\to K^-\pi^+\pi^+\pi^0$ can provide further insights into more complicated dynamics and substructures in the $D$-meson decays to two vector mesons $D \to VV$,
which have attracted a great deal of attention in both theory and experiment~\cite{VV1,VV2,VV3,VV4,PLB684-137,arXiv2303,JHEP242,JHEP09077}, but where there is limited available experimental information. 
Fig.~\ref{topology1} shows the leading Feynman diagrams of the two vector mesons decay $D^+ \to \bar{K}^{*}(892)^0\rho(770)^+$.
Furthermore, measurements of the $D$ meson decays to axial-vector and pseudoscalar mesons $D \to AP$, such as $D^{+} \to \bar{K}_1(1270)^0\pi^{+}$ and $D^{+} \to \bar{K}_1(1400)^0\pi^{+}$, are also beneficial for the understanding of the mixing angle between axial-vectors $\theta_{K_1}$~\cite{CHY:2012ggx,PRD125}.  

In this paper we present the amplitude analysis and BF measurement of the decay $D^{+} \to K^-\pi^{+}\pi^{+}\pi^{0}$, utilizing 7.93 fb$^{-1}$ of $e^+e^-$ collision data collected at a center-of-mass energy $\sqrt{s} = $ 3.773 GeV by the BESIII detector at BEPCII~\cite{lum1,lum2}. A double tag~(DT) method is employed by reconstructing $D^+D^-$ from $D^+ \to K^-\pi^+\pi^+\pi^0$ and $D^- \to K^+\pi^-\pi^-$, respectively. 
Charged-conjugate modes and exchange symmetry of two identical $\pi^+$ are always implied throughout.
\section{Detector and data}
\label{sec:detector_dataset}
The BESIII detector~\cite{Ablikim:2009aa} records symmetric $e^+e^-$ collisions 
provided by the BEPCII storage ring~\cite{Yu:IPAC2016-TUYA01}
in the center-of-mass energy range from 1.85 to 4.95~GeV, with a peak luminosity of $1.1 \times 10^{33}\;\text{cm}^{-2}\text{s}^{-1}$ 
achieved at $\sqrt{s} = 3.773\;\text{GeV}$. 
BESIII has collected large data samples in this energy region~\cite{Ablikim:2019hff, EcmsMea, EventFilter}. The cylindrical core of the BESIII detector covers 93\% of the full solid angle and consists of a helium-based
 multilayer drift chamber~(MDC), a plastic scintillator time-of-flight
system~(TOF), and a CsI(Tl) electromagnetic calorimeter~(EMC),
which are all enclosed in a superconducting solenoidal magnet
providing a 1.0~T magnetic field.
The solenoid is supported by an
octagonal flux-return yoke with resistive plate counter muon
identification modules interleaved with steel. 
The charged-particle momentum resolution at $1~{\rm GeV}/c$ is
$0.5\%$, and the ionization energy loss 
(${\rm d}E/{\rm d}x$)
resolution in MDC is $6\%$ for electrons
from Bhabha scattering. The EMC measures photon energies with a
resolution of $2.5\%$ ($5\%$) at $1$~GeV in the barrel (end-cap)
region. The time resolution in the TOF barrel region is 68~ps, while
that in the end-cap region was 110~ps.  The end-cap TOF
system was upgraded in 2015 using multigap resistive plate chamber
technology, providing a time resolution of
60~ps, which benefits 63\% of the data used in this analysis~\cite{etof,etof2,etof3}.

Monte Carlo (MC) simulated data samples produced with a {\sc
geant4}-based~\cite{geant4} software package, which
includes the geometric description of the BESIII detector and the
detector response, are used to determine detection efficiencies
and estimate backgrounds. The simulation models the beam-energy spread and initial-state radiation (ISR) in the $e^+e^-$
annihilations with the generator {\sc
kkmc}~\cite{ref:kkmc,ref:kkmc2}. The inclusive MC sample includes the production of $D\bar{D}$
pairs which contains quantum coherence for the neutral $D$ channels,
the non-$D\bar{D}$ decays of the $\psi(3770)$, the ISR
production of the $J/\psi$ and $\psi(3686)$ states, and the
continuum processes incorporated in {\sc kkmc}~\cite{ref:kkmc,ref:kkmc2}.
All particle decays are modeled with {\sc
evtgen}~\cite{ref:evtgen,ref:evtgen2} using BFs 
either taken from the
Particle Data Group~\cite{PDG}, when available,
or otherwise estimated with {\sc lundcharm}~\cite{ref:lundcharm,ref:lundcharm2}.
Final-state radiation
from charged final-state particles is incorporated using the {\sc
photos} package~\cite{photos2}.

In this work, two sets of MC samples are used: the phase space (PHSP) MC sample and the signal MC sample. For the tag process $D^- \to K^+\pi^+\pi^-$, both sets of MC samples are simulated using the model derived from the amplitude analysis reported in ref.~\cite{PRD052001}.
In the PHSP MC sample the signal process $D^+ \to K^-\pi^+\pi^+\pi^0$ is generated with a uniform distribution in PHSP in order to allow the calculation of  the normalization factor of the probability density function (PDF) used in the amplitude analysis. In the signal MC sample, the signal process is generated based on the results of the amplitude analysis and is used to estimate the detection efficiencies.

\section{Event selection}
\label{ST-selection} 

Charged tracks detected in the MDC are required to be within a polar angle ($\theta$) range of $|\rm{cos\theta}|<0.93$, where $\theta$ is defined with respect to the $z$-axis, which is the symmetry axis of the MDC. For charged tracks, the distance of closest approach to the interaction point (IP) must be less than 10\,cm along the $z$-axis, $|V_{z}|$,  and less than 1\,cm in the transverse plane, $|V_{xy}|$.

Photon candidates are identified using isolated showers in the EMC.  The deposited energy of each shower must be more than 25~MeV in the barrel region ($|\!\cos \theta|< 0.80$) and more than 50~MeV in the end-cap region ($0.86 <|\!\cos \theta|< 0.92$).  To exclude showers that originate from charged tracks, the angle subtended by the EMC shower and the position of the closest charged track at the EMC must be greater than 10 degrees as measured from the IP. To suppress electronic noise and showers unrelated to the event, the difference between the EMC time and the event start time is required to be within [0, 700]\,ns.

Particle identification~(PID) for charged tracks combines measurements of the d$E$/d$x$ and the flight time in the TOF to form likelihoods $\mathcal{L}(h)~(h=K,\pi)$ for each hadron $h$ hypothesis.
Charged kaons and pions are identified by comparing the likelihoods, $\mathcal{L}(K)>\mathcal{L}(\pi)$ and $\mathcal{L}(\pi)>\mathcal{L}(K)$, respectively.

The $\pi^0$ candidates are formed from the photon pairs with invariant masses in a range of $[0.115, 0.150]$~GeV/$c^{2}$, which is about three times the mass resolution. Moreover, in order to achieve an adequate resolution, at least one of the two photons is required 
to be detected in the barrel EMC. A one-constraint
 kinematic fit that constrains 
the $\gamma\gamma$ invariant mass to the known $\pi^{0}$ mass~\cite{PDG} is performed to improve the mass resolution. The $\chi^2$ of the kinematic fit is required
to be less than 30.

The $D^+D^-$ pairs are reconstructed from $D^+ \to K^-\pi^+\pi^+\pi^0$ and $D^- \to K^+\pi^-\pi^-$, respectively. To distinguish the $D^+D^-$ mesons from the backgrounds, the beam-constrained mass~($M_{\rm{BC}}$) and the energy difference~($\Delta E$) are used to identify the signal $D^+D^-$ pair:
\begin{equation}
\begin{aligned}
	&M_{\rm{BC}} = \sqrt{E_{\rm{beam}}^2-|\vec{p}_{D}|^2}, \\
	&\Delta E = E_{D} - E_{\rm{beam}},
\end{aligned}
\end{equation}
where $\vec{p}_{D}$ and $E_{D}$ are the total reconstructed momentum and energy of the $D$ candidate, and $E_{\rm{beam}}$ is the beam energy. The $D$ signal manifest itself as a peak around the known $D$ mass~\cite{PDG} in the $M_{\rm{BC}}$ distribution and as a peak around zero in the $\Delta E$ distribution. If multiple DT candidates are present in an event, the one with the smallest quadratic sum of $\Delta E$ from the signal and tag sides ($\Delta E_{\rm sig}^2+\Delta E_{\rm tag}^2$ ) is selected for further analysis.

\section{Amplitude analysis}
\label{Amplitude-Analysis}
\subsection{Further selection criteria}
\label{sec:pwa-select}

To increase the signal purity for the amplitude analysis, the requirement of $-0.062 < \Delta E < 0.034$~($-0.025 < \Delta E < 0.025$) GeV for $D^+ \to K^-\pi^+\pi^+\pi^0$~($D^- \to K^+\pi^-\pi^-$) is applied. 
To suppress background from 
$D^0 \to K^-\pi^+\pi^+\pi^-$, $\bar{D}^0 \to K^+\pi^-\pi^{\rm{0}}$ events in the $D^+ \to K^-\pi^+\pi^+\pi^{\rm{0}}$, $D^- \to K^+\pi^-\pi^-$ sample, where  the $\pi^{\rm{0}}$ from the $\bar{D}^0$ decay and the $\pi^-$ from the $D^0$ decay are interchanged, we reconstruct the wrong beam-constrained mass~($M_{\rm{BC}}^{\rm W}$) and the wrong energy difference~($\Delta E^{\rm W}$) according to the $D^0\bar{D}^0$ decay mode hypothesis. For multiple misidentified candidates, we use the minimum quadratic sum of $\Delta E^{\rm W}$ to select the “best background” event. The $D^0\bar{D}^0$ backgrounds form a peak around the known $D^0$ mass~\cite{PDG} while the distribution for signal is flat.  Events satisfying 1.863 $< M_{\rm{BC}}^{\rm{W}} <$ 1.867 GeV/$c^2$ for both tag and signal sides are rejected. 

A six-constraint kinematic fit is performed, in which the four-momenta of the final-state particles are constrained to the initial four-momenta of the $e^+e^-$ system and the reconstructed masses of $D^+$ and $\pi^{\rm{0}}$ are constrained to their known values~\cite{PDG}. The events with $\chi^2<100$ are retained, and the modified four-momenta of the final state particles from the kinematic fit are used to perform the amplitude analysis. 

\begin{figure}[htbp]
  \centering
 \includegraphics[width=0.45\textwidth]{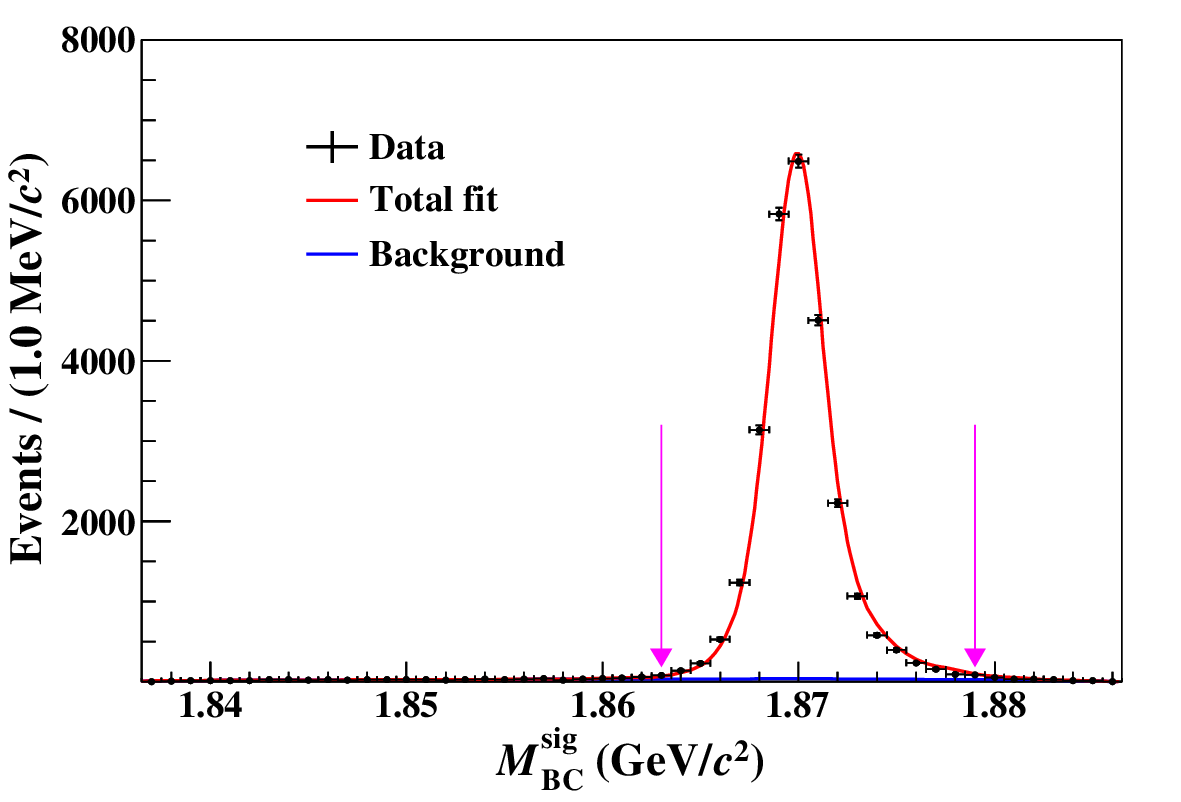}
 \includegraphics[width=0.45\textwidth]{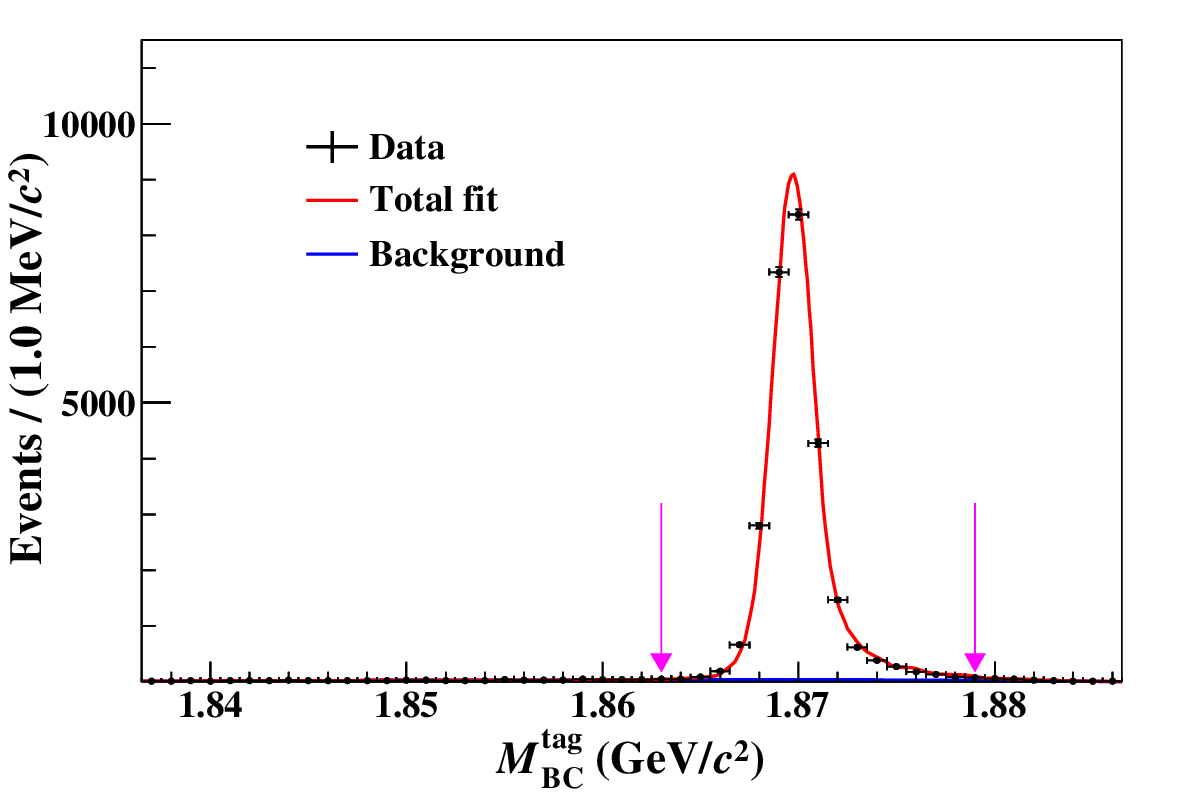}~\\
 \includegraphics[width=0.45\textwidth]{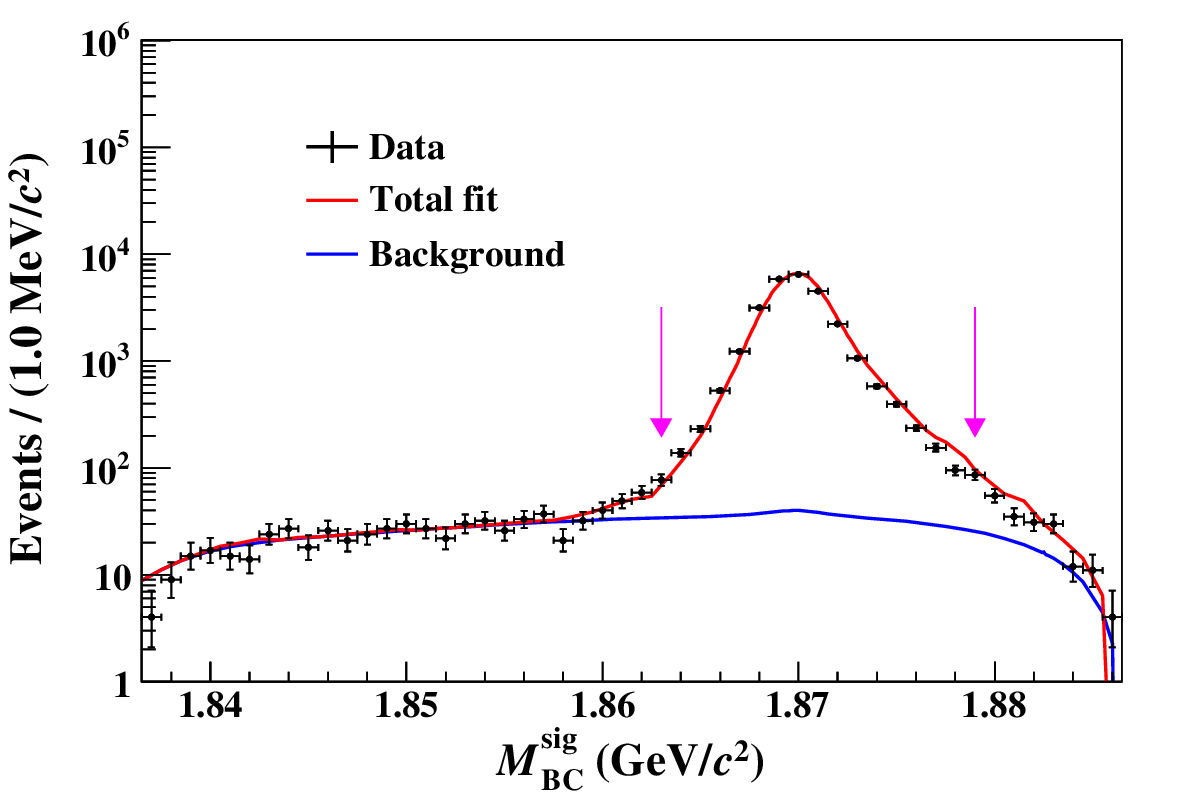}
 \includegraphics[width=0.45\textwidth]{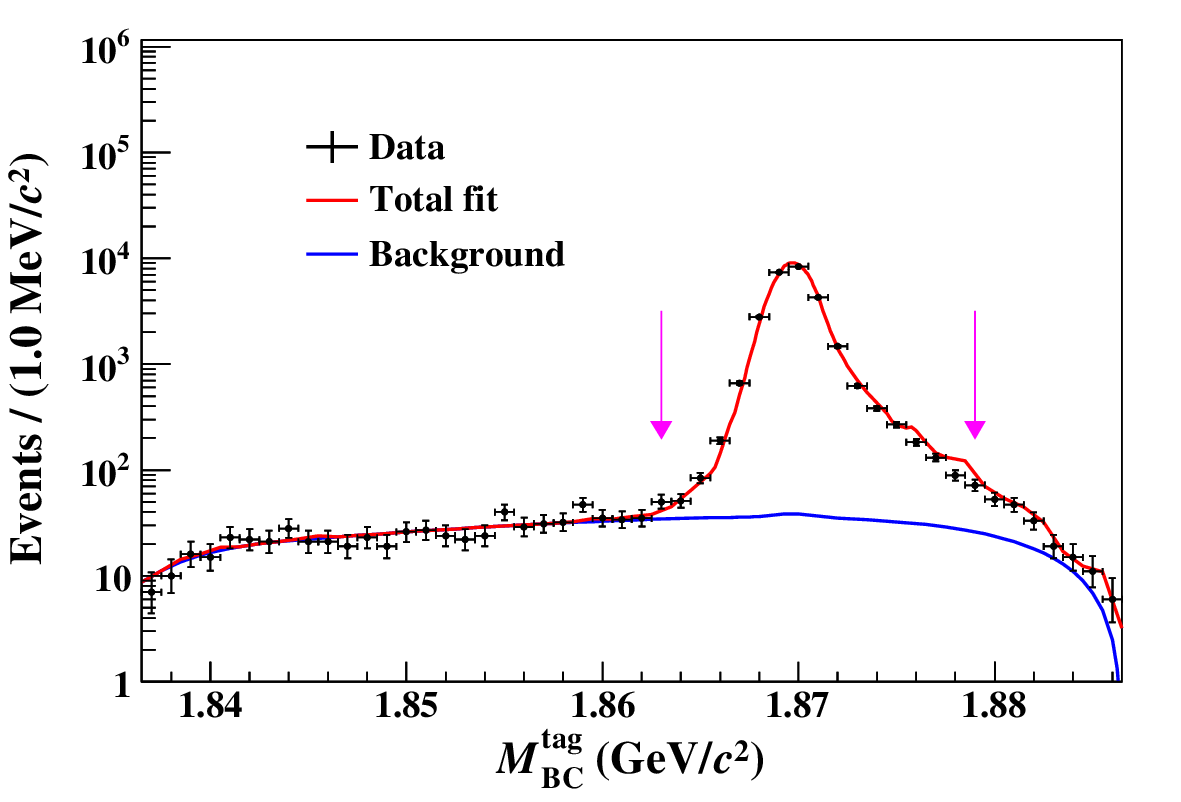}
  \caption{Projections on the $M_{\rm{BC}}$ distributions of the 2D fit described in the text for the signal~(left) and tag~(right) sides. The distribution corresponding to the Y-axis logarithmic coordinates is listed in the following two figures. 
The arrows indicated the boundaries of the selected signal region.}
  \label{fig:2dfit}
\end{figure}

After applying all of the aforementioned criteria, the signal yields are extracted from an unbinned two-dimensional (2D) maximum likelihood fit to the distribution of $M_{\rm{BC}}^{\rm{sig}}$ versus $M_{\rm{BC}}^{\rm{tag}}$~(see Appendix~\ref{2dfit} for details). 
The fit results are depicted in Fig.~\ref{fig:2dfit}. A total of 26,709 events with a purity~($P_{\mathcal{S}}$) of $(98.4\pm 0.1)\%$ in both the $M_{\rm{BC}}^{\rm{sig}}$ and $M_{\rm{BC}}^{\rm{tag}}$ signal region 
of [1.863, 1.879] GeV/$c^2$ are retained for the subsequent amplitude analysis.

\subsection{Fit method}
\label{sec:fitmethod}

An unbinned maximum-likelihood fit is used in the amplitude analysis of $D^+ \to K^-\pi^+\pi^+\pi^0$. The likelihood function $\mathcal{L}$ is constructed with the signal and background PDFs, which depend on the momenta of the four final-state particles.
Given the high purity of the signal events, the log-likelihood function is constructed by summing all signal candidates and subtracting the MC-simulated backgrounds:\begin{eqnarray}\begin{aligned}
 \ln~\mathcal{L} = \sum_{k_{\rm{data}}=1}^{N_{\rm{data}}}\ln~f_{\rm data}(p^{k_{\rm{data}}})-\sum_{k_{\rm{bkg}}=1}^{N_{\rm{bkg}}}w_{\rm bkg}\ln~f_{\rm bkg}(p^{k_{\rm bkg}})\,,  \label{likelihood3}
\end{aligned}\end{eqnarray}
where $k_{\rm{data}}$ and $N_{\rm{data}}$ represent the index of the $k^{\rm th}$ event and the number of the selected signal candidates, respectively. The notation $p^{k_{\rm data}}$ is used to describe the four-momenta of the final-state particles for the $k^{\rm th}$ data candidate, and $f_{\rm{data}}$ represents the PDF of the data candidate.
The symbols with the subscript ``bkg'' represent the corresponding parameters associated with the MC-simulated backgrounds. Furthermore,  $w_{\rm bkg}$ is the weight of MC-simulated backgrounds that is determined by $N_{\rm{data}} \times (1-P_{\mathcal{S}})/N_{\rm{bkg}}$, where $P_{\mathcal{S}}$ is the signal purity discussed in Sec.~\ref{sec:pwa-select}.

The PDF for the data candidates and MC-simulated backgrounds are both given by 
\begin{eqnarray}\begin{aligned}
  f(p) = \frac{\epsilon(p)\left|\mathcal{M}(p)\right|^{2}R_{4}}{\int \epsilon(p)\left|\mathcal{M}(p)\right|^{2}R_{4}\,{\rm d}p}\,, \label{signal-PDF}
\end{aligned}\end{eqnarray}
where $\epsilon(p)$ is the detection efficiency parameterized in terms of the four-momenta $p$, and $R_{4}$ is the PHSP factor for four-body decays. 
The total amplitude $\mathcal{M}$ is modeled with the isobar model, which is the coherent sum of the individual amplitudes of intermediate processes and is given by $\mathcal{M}=\sum \rho_{n}e^{i\phi_{n}}\mathcal{A}_{n}$, where the magnitude $\rho_{n}$ and phase $\phi_{n}$ are the free parameters to be determined by the fit. The amplitude of the $n^{\rm th}$ intermediate process~($\mathcal{A}_{n}$) is given by
\begin{eqnarray}
\begin{aligned}
  \mathcal{A}_{n} = P_n^1P_n^2S_nF_n^1F_n^2F_n^{3}\,, \label{base-amplitude}
\end{aligned}\end{eqnarray}
where the indices 1, 2 and 3 correspond to the two subsequent intermediate resonances and the $D^+$ meson, respectively. Here, 
$F_{n}$ is the Blatt-Weisskopf barrier (Sec.~\ref{sec:barrier}),
$P_{n}$ is the propagator of the intermediate resonance (Sec.~\ref{sec:propagator}), and
$S_{n}$ is the spin factor constructed with the covariant tensor formalism~\cite{covariant-tensors} (Sec.~\ref{sec:spinfactor}). 
The normalization integral is realized by a signal MC integration,
\begin{eqnarray}\begin{aligned}
  \int \epsilon(p) |\mathcal{M}(p)|^2 R_{4}\,{\rm d}p \approx
\frac{1}{N_{\rm MC}}\sum_{k_{\rm MC}}^{N_{\rm MC}} \frac{ |\mathcal{M}(p^{k_{\rm MC}})|^2 }{\left|\mathcal{M}^{g}(p^{k_{\rm MC}})\right|^{2}}\,, \label{MC-intergral}
\end{aligned}\end{eqnarray}
where $k_{\rm MC}$ is the index of the $k^{\rm th}$ event of the signal MC sample, and $N_{\rm MC}$ is the number of the selected signal MC events.
The symbol $\mathcal{M}^{g}(p)$ denotes the PDF used to generate the signal MC sample in the MC integration.

To account for the bias caused by differences in tracking, PID efficiencies, and $\pi^0$ reconstruction between data and MC simulation, each signal MC event is weighted with a ratio, $\gamma_{\epsilon}(p)$, which is calculated as
\begin{equation}
	\gamma_{\epsilon}(p) = \prod_{j} \frac{\epsilon_{j,\rm data}(p)}{\epsilon_{j,\rm MC}(p)},
	\label{pwa:gamma}
\end{equation}
where $j$ denotes the four final-state particles, $\epsilon_{j,\rm data}(p)$ and $\epsilon_{j,\rm MC}(p)$ are the tracking, PID and $\pi^0$ reconstruction efficiencies as a function of the momentum of the final-state particles for data and MC simulation, respectively.
Then the MC integration is determined by 
\begin{eqnarray}\begin{aligned}
    &\int \epsilon(p) |\mathcal{M}(p)|^2 R_{4}\,{\rm d}p \approx
&\frac{1}{N_{\rm MC}} \sum_{k_{\rm MC}}^{N_{\rm MC}} \frac{ |\mathcal{M}(p^{k_{\rm MC}})|^2 \gamma_{\epsilon}(p^{k_{\rm MC}})}{\left|\mathcal{M}^{g}(p^{k_{\rm MC}})\right|^{2}}\,.
\label{MC-intergral-corrected}
\end{aligned}\end{eqnarray}

\subsubsection{Blatt-Weisskopf barrier factors}\label{sec:barrier}
For the process $a \to bc$, the Blatt-Weisskopf barrier factors~\cite{BW}, 
$X_L(q)$, are
parameterized as a function of the angular momentum $L$ and the momenta $q$ of
the final-state particle $b$ or $c$ in the rest system of $a$. They are taken as
\begin{eqnarray}
\begin{aligned}
 X_{L=0}(q)&=1,\\
 X_{L=1}(q)&=\sqrt{\frac{z_0^2+1}{z^2+1}},\\
 X_{L=2}(q)&=\sqrt{\frac{z_0^4+3z_0^2+9}{z^4+3z^2+9}}\,,
\end{aligned}
\end{eqnarray}
where $z=qR_r$, $z_0=q_0R_r$ and the effective radius, $R_r$, of the barrier is fixed to 3.0~GeV$^{-1}$ for the intermediate 
resonances and 5.0~GeV$^{-1}$ for the $D^+$ meson. 
The momentum $q$ is given by
\begin{eqnarray}
\begin{aligned}
q = \sqrt{\frac{(s_a+s_b-s_c)^2}{4s_a}-s_b}\,, \label{q2}
\end{aligned}
\end{eqnarray}
where the value of $q_0$ is that of $q$ when $s_a$ takes squared of the rest mass of particle $a$, and $s_a(s_b,s_c)$ denotes the squared invariant-mass of the system consisting of $a(b,c)$.
\subsubsection{Propagator}\label{sec:propagator}
The intermediate resonances $\bar{K}^{*}(892)^0$, $\bar{K}_1(1270)^0$, $\bar{K}_1(1400)^0$, $\bar{K}(1460)^0$, and $\bar{K}^*(1680)^0$ are
parameterized with the relativistic Breit-Wigner~(RBW) function,
\begin{eqnarray}\begin{aligned}
    P(m) = \frac{1}{m_{0}^{2} - m^2 - im_{0}\Gamma(m)}\,,\; 
    \Gamma(m) = \Gamma_{0}\left(\frac{q}{q_{0}}\right)^{2L+1}\left(\frac{m_{0}}{m}\right)X^{2}_{L}(q)\,, 
  \label{RBW}
\end{aligned}\end{eqnarray}
where $m$ is the invariant mass of the decay products, $m_0$ and $\Gamma_0$ are the mass and width of the intermediate resonance that are fixed to their known values~\cite{PDG}. The energy-dependent width is denoted by $\Gamma(m)$.

The decay of the $\bar{K}_1(1270)^0 \to K^-\pi^+\pi^0$ proceeds through a quasi-three-body process, with a complex energy-dependent width that lacks a general analytic expression. The corresponding values are obtained through an iterative method of integrating the squared transition amplitude over the three-body PHSP~\cite{K1270}.

The $\rho(770)^+$ resonance is parameterized by the
Gounaris-Sakurai~(GS) lineshape~\cite{GS}, which is given by
\begin{eqnarray}\begin{aligned}
P_{\rm GS}(m)=\frac{1+\mathcal{C}_{\rm GS}\frac{\Gamma_0}{m_0}}{m_0^2-m^2+f(m)-im_0\Gamma(m)}\,.
\end{aligned}\end{eqnarray}
The normalization condition at $P_{\rm GS}(0)$ fixes the parameter
$\mathcal{C}_{\rm GS}=f(0)/(\Gamma_0 m_0)$. It is found to be
\begin{eqnarray}\begin{aligned}
\mathcal{C}_{\rm GS}=\frac{3m^2_\pi}{\pi q_0^2}\ln\left(\frac{m_0+2q_0}{2m_\pi}\right)+\frac{m_0}{2\pi q_0}-\frac{m^2_\pi m_0}{\pi q^3_0}\,,
\end{aligned}\end{eqnarray}
where $m_{\pi}$ is the known $\pi$ mass~\cite{PDG}, . The function $f(m)$ is given by
\begin{eqnarray}
\begin{aligned}
f(m)=\Gamma_0\frac{m_0^2}{q_0^3}\left[q^2(h(m)-h(m_0))+(m_0^2-m^2)q_0^2\left.\frac{dh(m)}{d(m^2)}\right|_{m_0^2}\right]\,,
\end{aligned}
\end{eqnarray}
where
\begin{eqnarray}\begin{aligned}
h(m)=\frac{2q}{\pi m}\ln\left(\frac{m+2q}{2m_{\pi}}\right)\,.
\end{aligned}\end{eqnarray}
And the derivative of $h(m)$ with respect to $m^2$ evaluated at $m_0^2$ is calculated by
\begin{eqnarray}\begin{aligned}
&\left.\frac{dh(m)}{d(m^2)}\right|_{m_0^2}=
&h(m_0)\left[(8q_0^2)^{-1}-(2m_0^2)^{-1}\right]+(2\pi m_0^2)^{-1}\,.
\end{aligned}\end{eqnarray}

The $K\pi$ $S$-wave is modeled by the LASS parameterization~\cite{KPsnew}, which is described by a $K_0^{*}(1430)$ Breit-Wigner together with an effective range non-resonant component with a phase shift. It is given by	
\begin{equation}
A(m)=F\sin\delta_Fe^{i\delta_F}+R\sin\delta_Re^{i\delta_R}e^{i2\delta_F},
\end{equation}
with
\begin{equation}
\begin{aligned}
&\delta_F=\phi_F+\cot^{-1}\left[\frac{1}{aq}+\frac{rq}{2}\right], \\
&\delta_R=\phi_R+\tan^{-1}\left[\frac{M_{K^*_0(1430)}\Gamma(m_{K\pi})}{M_{K^*_0(1430)}^2-m^2_{K\pi}}\right].
\end{aligned}
\end{equation}
The parameters $F$, $\phi_F$~($R$ and $\phi_R$) are the amplitudes and phases of the non-resonant~(resonant) component, respectively. The parameters $a$ and $r$ are the scattering length and effective interaction length, respectively. 
The parameters $M_{K^*_0(1430)}$ and $m_{K\pi}$ are the defined $K^*_0(1430)$ mass and the invariant mass of  the $K\pi$ system, respectively.
We fix these parameters $(M_{K^*_0(1430)}, \Gamma, F, \phi_F , R, \phi_R, a, r)$ to the results obtained from the amplitude analysis of a sample of $D^0\to K_S^0\pi^+\pi^-$ decays by the BaBar and Belle experiments~\cite{KPsnew2}. The values of these parameters are summarized in Table~\ref{tab:babar}. 
\begin{table}[h]
\setlength{\abovecaptionskip}{0.cm}
\setlength{\belowcaptionskip}{-0.2cm}
  \begin{center}
    \begin{tabular}{|lc|}
			\hline
      $M_{K^*_0(1430)}$(GeV/$c^2$)        &1.441 $\pm$ 0.002\\
      $\Gamma$(GeV)   &0.193 $\pm$ 0.004\\
      $F$                 &0.96 $\pm$ 0.07\\
      $\phi_F$~($^\circ$)            &0.1 $\pm$ 0.3\\
      $R$                 &1(fixed)\\
      $\phi_R$~($^\circ$)            &$-$109.7 $\pm$ 2.6\\
      $a$ (GeV/$c$)$^{-1}$                &0.113 $\pm$ 0.006\\
      $r$ (GeV/$c$)$^{-1}$                  &$-$33.8 $\pm$ 1.8\\
      \hline
    \end{tabular}
  \end{center}
  \caption{The $K\pi$ $S$-wave parameters are obtained from the amplitude analysis of $D^0\to K_S^0\pi^+\pi^-$ in the BaBar and Belle experiments~\cite{KPsnew2}. The uncertainties are the combined statistical and systematic uncertainties.}
    \label{tab:babar}

\end{table}
\subsubsection{Spin factors}\label{sec:spinfactor}
Due to the limited size of the PHSP, we only consider states with angular momenta below three units. For the process $a \to bc$, the four momenta of the particles $a$, $b$, and $c$ are denoted as $p_a$, $p_b$, and $p_c$, respectively.
The spin-projection operators~\cite{covariant-tensors} are defined as
\begin{eqnarray}
\begin{aligned}
  &P^{(0)}(a) = 1\,,&(S~\rm{wave})\\
  &P^{(1)}_{\mu\mu^{\prime}}(a) = -g_{\mu\mu^{\prime}}+\frac{p_{a,\mu}p_{a,\mu^{\prime}}}{p_{a}^{2}}\,,&(P~\rm{wave})\\
  &P^{(2)}_{\mu\nu\mu^{\prime}\nu^{\prime}}(a) = \frac{1}{2}(P^{(1)}_{\mu\mu^{\prime}}(a)P^{(1)}_{\nu\nu^{\prime}}(a)+P^{(1)}_{\mu\nu^{\prime}}(a)P^{(1)}_{\nu\mu^{\prime}}(a)) -\frac{1}{3}P^{(1)}_{\mu\nu}(a)P^{(1)}_{\mu^{\prime}\nu^{\prime}}(a)\,.&(D~\rm{wave})
 \label{spin-projection-operators}
\end{aligned}
\end{eqnarray}
The pure orbital angular-momentum covariant tensors are given by 
\begin{eqnarray}
\begin{aligned}
    \tilde{t}^{(0)}_{\mu}(a) &= 1\,,&(S~\rm{wave})\\
    \tilde{t}^{(1)}_{\mu}(a) &= -P^{(1)}_{\mu\mu^{\prime}}(a)r^{\mu^{\prime}}_{a}\,,&(P~\rm{wave})\\
    \tilde{t}^{(2)}_{\mu\nu}(a) &= P^{(2)}_{\mu\nu\mu^{\prime}\nu^{\prime}}(a)r^{\mu^{\prime}}_{a}r^{\nu^{\prime}}_{a}\,,&(D~\rm{wave})\\
\label{covariant-tensors}
\end{aligned}
\end{eqnarray}
where $r_a = p_b-p_c$. The spin factors $S(p)$ for the various components used in the analysis are listed in Table~\ref{table:spin_factors}. The tensor describing the $D^+$ decays with orbital angular-momentum quantum number $l$ is denoted by $\tilde{T}^{(l)\mu}$ and that of the intermediate $a \to bc$ decay is denoted by $\tilde{t}^{(l)\mu}$, and the $\tilde{T}^{(l)\mu}$ has the same definition as $\tilde{t}^{(l)\mu}$ in Ref.~\cite{covariant-tensors}.
\begin{table}[h]
\setlength{\abovecaptionskip}{0.cm}
\setlength{\belowcaptionskip}{0.cm}
 \begin{center}
\begin{tabular}{|lc|}
\hline
 Decay chain& $S(p)$ \\
 \hline
$D^+[S] \rightarrow V_1V_2$ & $P^{(1)\mu\nu}(D^+)\tilde{t}^{(1)\mu}(V_1) \; \tilde{t}^{(1)}_\nu(V_2)$ \\
$D^+[P] \rightarrow V_1V_2$ & $\epsilon_{\mu\nu\lambda\sigma}p^\mu(D^+) \; \tilde{T}^{(1)\nu}(D^+)\tilde{t}^{(1)\lambda}(V_1) \; \tilde{t}^{(1)\sigma}(V_2) $ \\
$D^+ \rightarrow AP_1, A[S] \rightarrow VP_2$ & $\tilde{T}^{(1)\mu}(D^+) \; P^{(1)}_{\mu\nu}(A) \; \tilde{t}^{(1)\nu}{(V)}$ \\
$D^+ \rightarrow AP_1, A[D] \rightarrow VP_2$ & $\tilde{T}^{(1)\mu}(D^+) \; \tilde{t}^{(2)}_{\mu\nu}(A) \; \tilde{t}^{(1)\nu}{(V)}$ \\
$D^+ \rightarrow V_1P_1, V_1 \rightarrow V_2P_2$ & $\epsilon_{\mu\nu\lambda\sigma}p^\mu_{V1}r^\nu_{V1}p^\lambda_{P1}r^\sigma_{V2}$ \\
$D^+ \rightarrow PP_1, P\rightarrow VP_2$ & $p^{\mu}(P_{2})\tilde{t}^{(1)}_{\mu}(V)$\\
$D^+ \rightarrow SV$ &$\tilde{T}^{(1)\mu}(D^+) \; \tilde{t}^{(1)}_{\mu}(V)$\\
\hline
\end{tabular}
\end{center}
 \caption{The spin factors $S(p)$ for the various contributions in the amplitude model. All operators, i.e.~$\tilde{t}$ and ~$\tilde{T}$, have the same definitions as in Ref.~\cite{covariant-tensors}.
  Scalar, pseudo-scalar, vector and axial-vector states are denoted
  by $S$, $P$, $V$ and $A$, respectively. The $[S]$ and $[P]$ denote the orbital angular-momentum quantum numbers $L = 0$  and 1, respectively.}\label{table:spin_factors}
\end{table}

\subsection{Fit results}
Using the method described in Sec.~\ref{sec:fitmethod}, we perform the fit in steps
by adding resonances one by one. The statistical significance of the newly added resonance is calculated by considering the change in the log likelihood value and taking into account the change in the number of degrees of freedom.

The data-fitting process commences with a base model incorporating the amplitudes of $D^{+}\to \bar{K}^{*}(892)^0\rho(770)^+$ and  $D^{+}\to \bar{K}_1(1400)^0\pi^+~(\bar{K}_1(1400)^0\to \bar{K}^{*}(892)\pi)$, as they are clearly observed in the corresponding invariant-mass spectra. The amplitude of $\bar{K}_1(1400)^0\to \bar{K}^{*}(892)\pi$ is a combination of the amplitudes of $\bar{K}^*(892)^-\pi^+$ and $\bar{K}^*(892)^0\pi^0$, taking into account the Clebsch-Gordan (CG) relation, which is detailed in Appendix~\ref{CGrelation}. The amplitudes $\bar{K}(1460)^0\to \bar{K}^{*}(892)\pi$ and $\bar{K}^*(1680)^0\to \bar{K}^{*}(892)\pi$ are also subject to the same relation. 

The amplitudes $D^{+}\to \bar{K}_1(1270)^0\pi^+$~$(\bar{K}_1(1270)^0\to K^- \rho(770)^+)$, $D^{+}\to\bar{K}(1460)^0\pi^+$
$(\bar{K}(1460)^0$$\to$ $\bar{K}^{*}(892)\pi)$, and $D^{+}\to \bar{K}^{*}(1680)^{0}\pi^+~(\bar{K}^{*}(1680)^{0}\to \bar{K}^{*}(892)\pi)$ are added and the change in the fit quality is assessed. As  $D^{+}\to (K^-\pi^+)_{S \rm{-wave}}\rho(770)^+$ and other non-resonance decay modes exhibit significances exceeding 5$\sigma$ and help improving the fit quality, they are also included in the model.
A comprehensive list of the other allowed contributions (based on known states) with statistical significances less than $5\sigma$ is provided in Appendix~\ref{tested_amplitude}. 

The fit fraction~(FF) for the $n^{\rm{th}}$ amplitude is computed numerically with generator-level MC events with a definition 
\begin{eqnarray}\begin{aligned}
  {\rm FF}_{n} = \frac{\sum^{N_{\rm gen}} \left|\rho_{n}e^{i\phi_{n}}\mathcal{A}_{n}\right|^{2}}{\sum^{N_{\rm gen}} \left|\mathcal{M}\right|^{2}}\,, \label{Fit-Fraction-Definition}
\end{aligned}\end{eqnarray}
where $N_{\rm gen}$ is the number of PHSP MC events at generator level.
The sum of these FFs may not be unity if there is net constructive or destructive
interference. Interference~(IN) between the $n^{\rm{th}}$ and $n^{\prime\rm{th}}$ amplitudes is defined as 
\begin{eqnarray}\begin{aligned}
  {\rm IN}_{nn^{\prime}} = \frac{\sum^{N_{\rm gen}} 2Re[\rho_{n}e^{i\phi_{n}}\mathcal{A}_{n}(\rho_{n^{\prime}}e^{i\phi_{n^{\prime}}}\mathcal{A}_{n^{\prime}})^{*}]}{\sum^{N_{\rm gen}} \left|\mathcal{M}\right|^{2}}\,. \label{interferenceFF-Definition}
\end{aligned}\end{eqnarray}
Here, the $Re$ in the numerator takes the modulus of that component. The interferences between the amplitudes are listed in Table~\ref{table:inter} of Appendix~\ref{app:interference}.

\begin{figure}
	        \centering
    \includegraphics[width=0.45\textwidth]{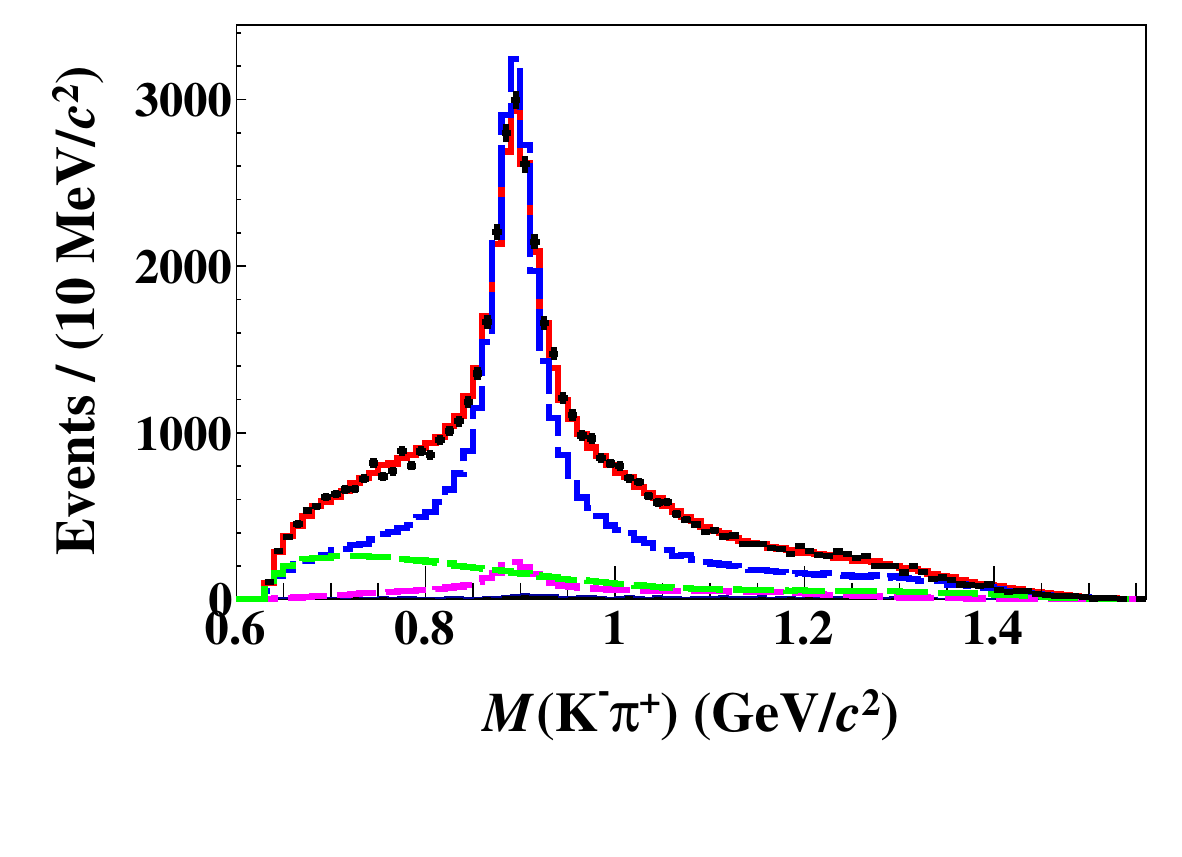}
    \includegraphics[width=0.45\textwidth]{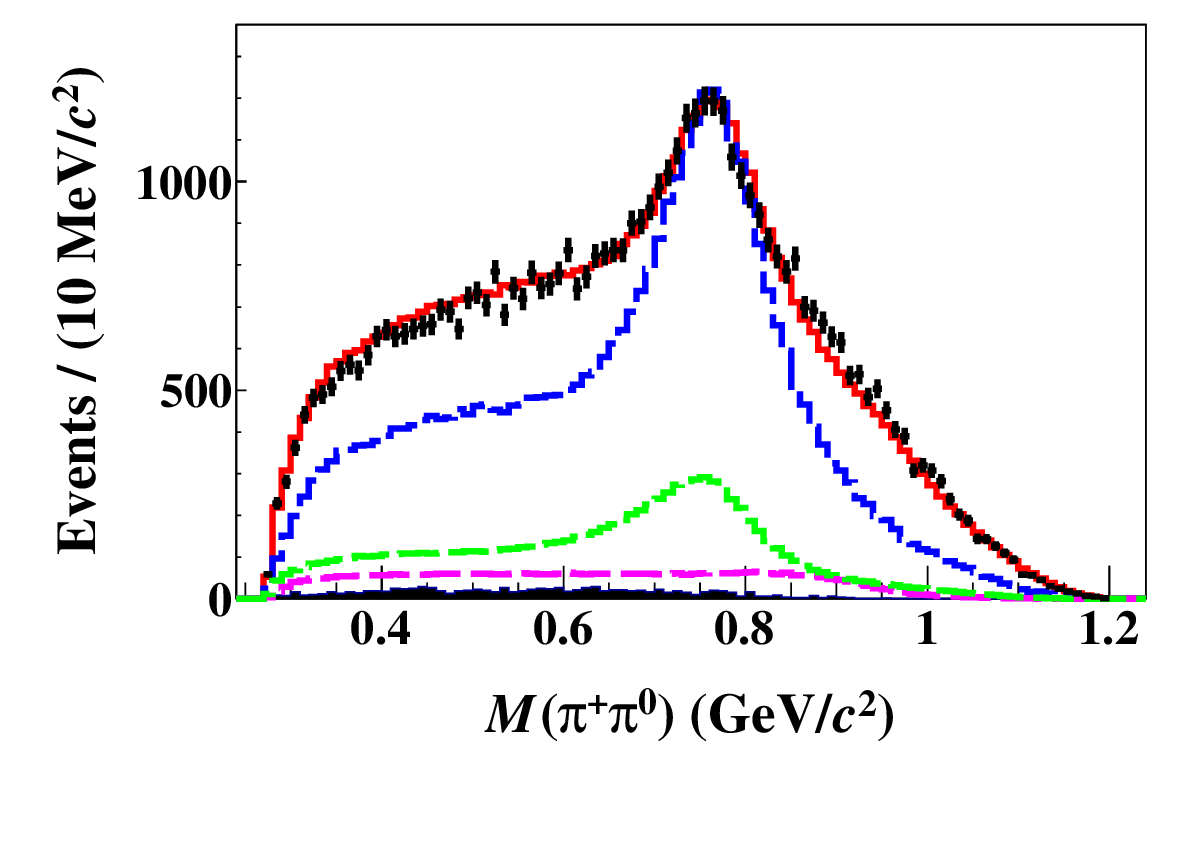}
    \includegraphics[width=0.45\textwidth]{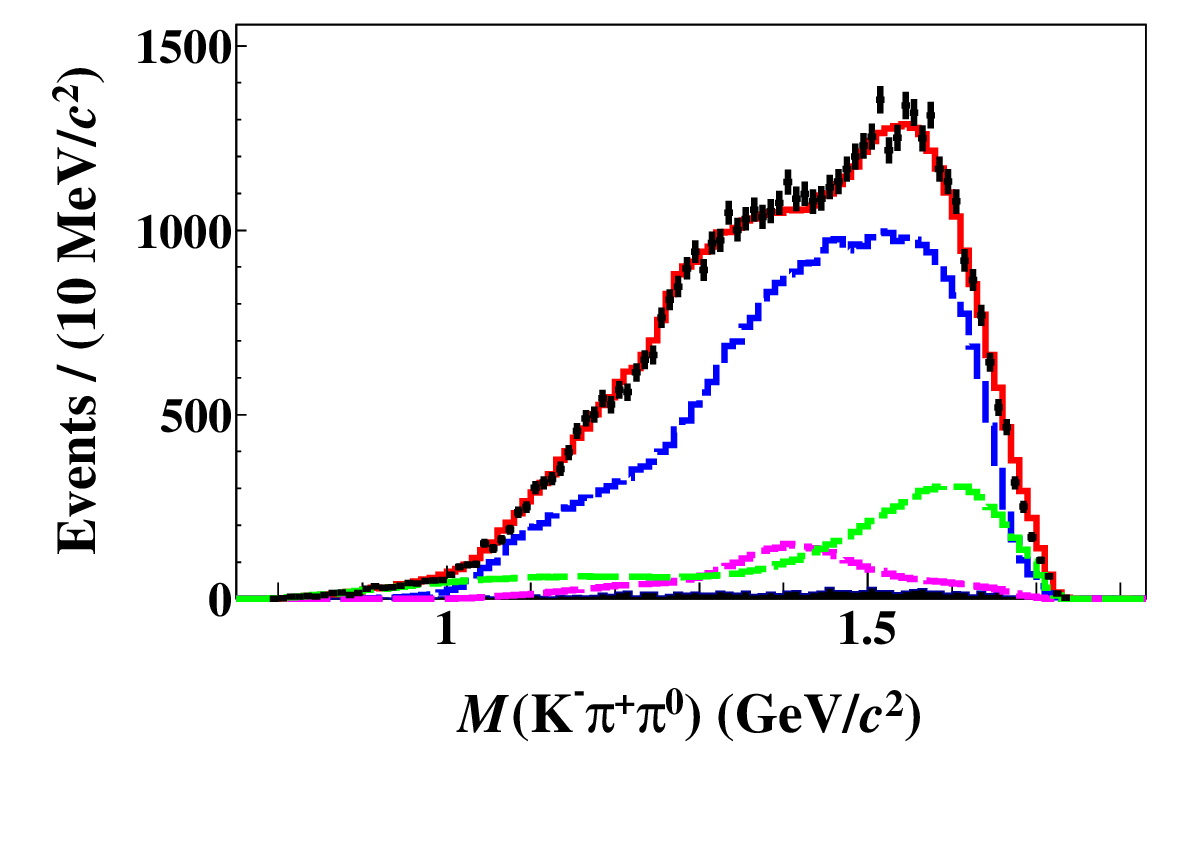}
    \includegraphics[width=0.45\textwidth]{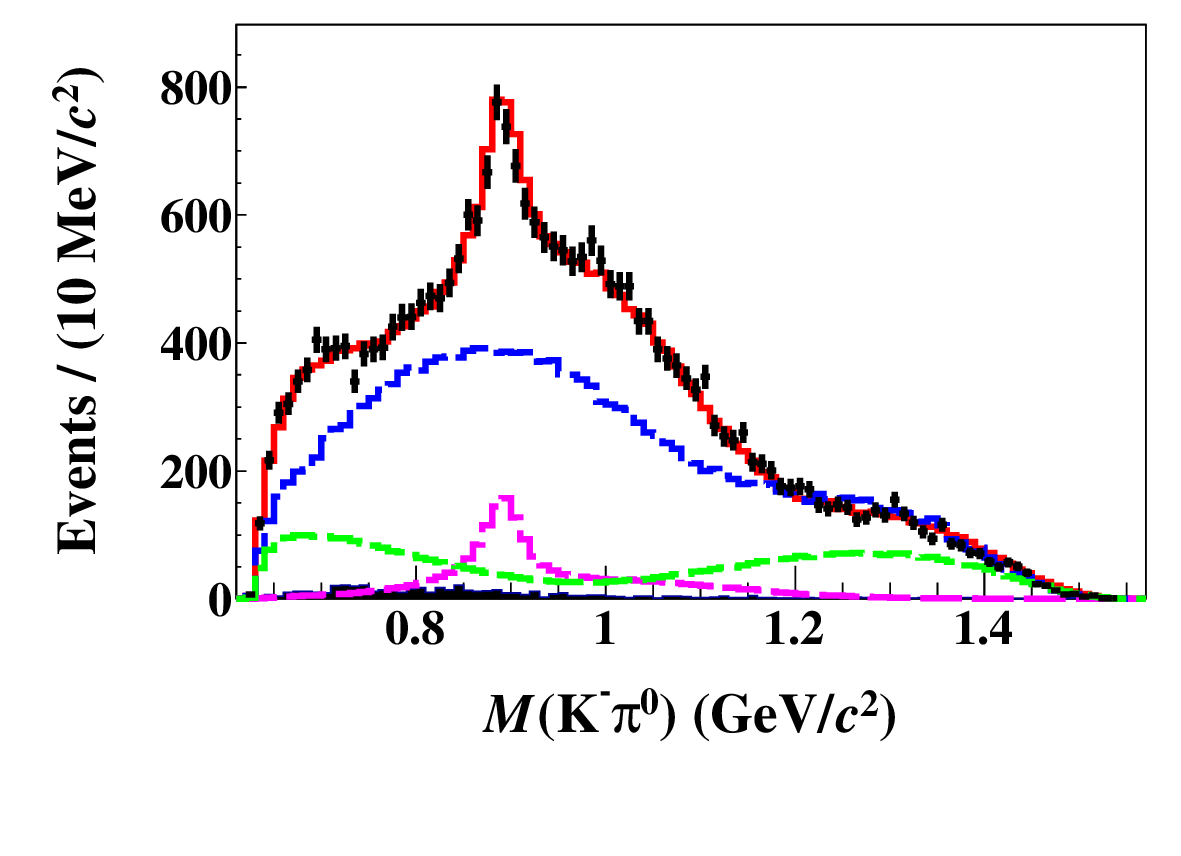}
    \includegraphics[width=0.45\textwidth]{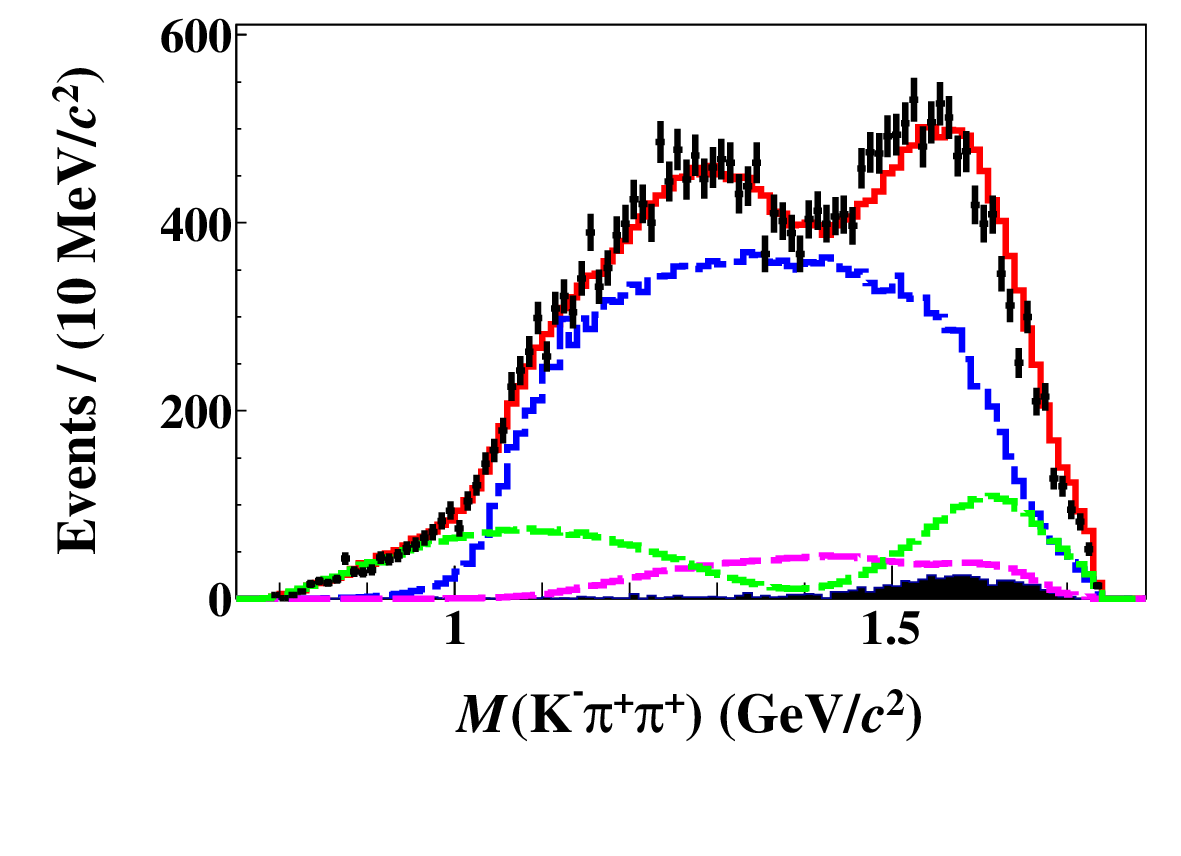}
     \includegraphics[width=0.45\textwidth]{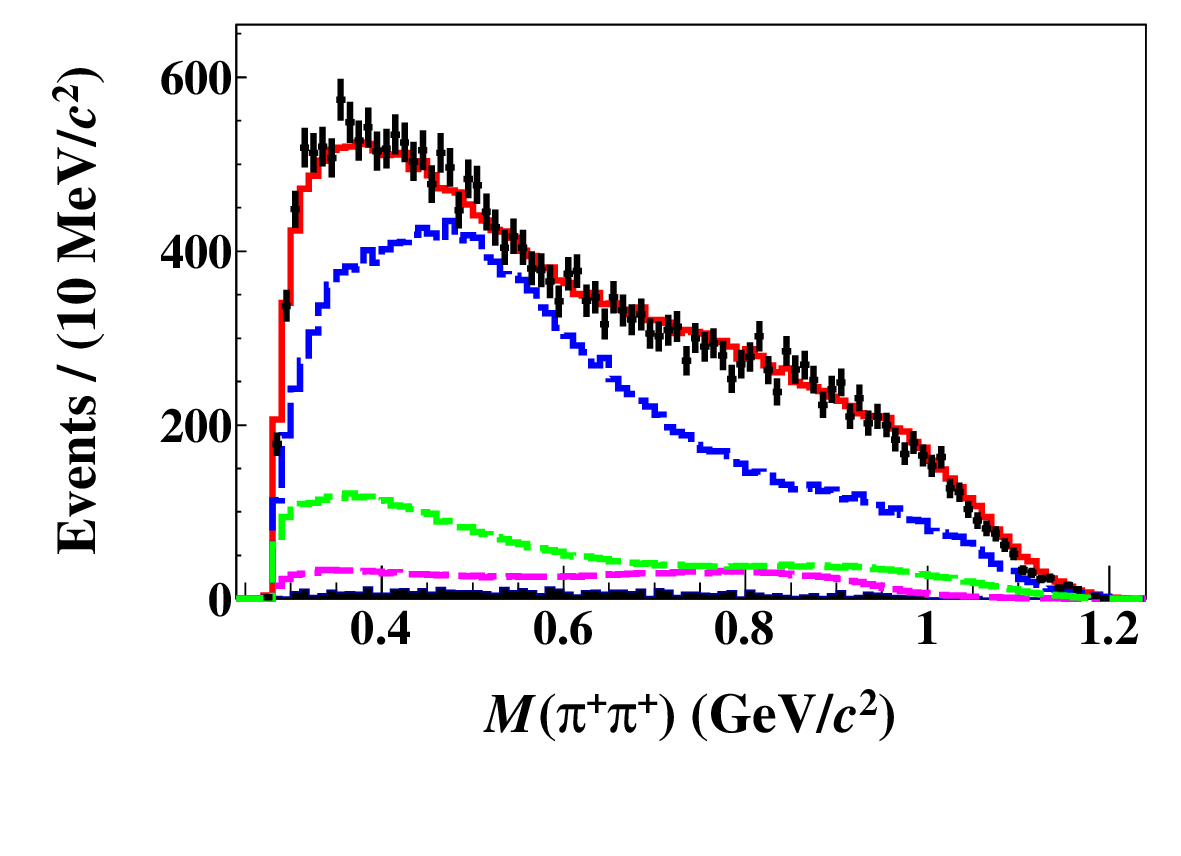}
     \includegraphics[width=0.45\textwidth]{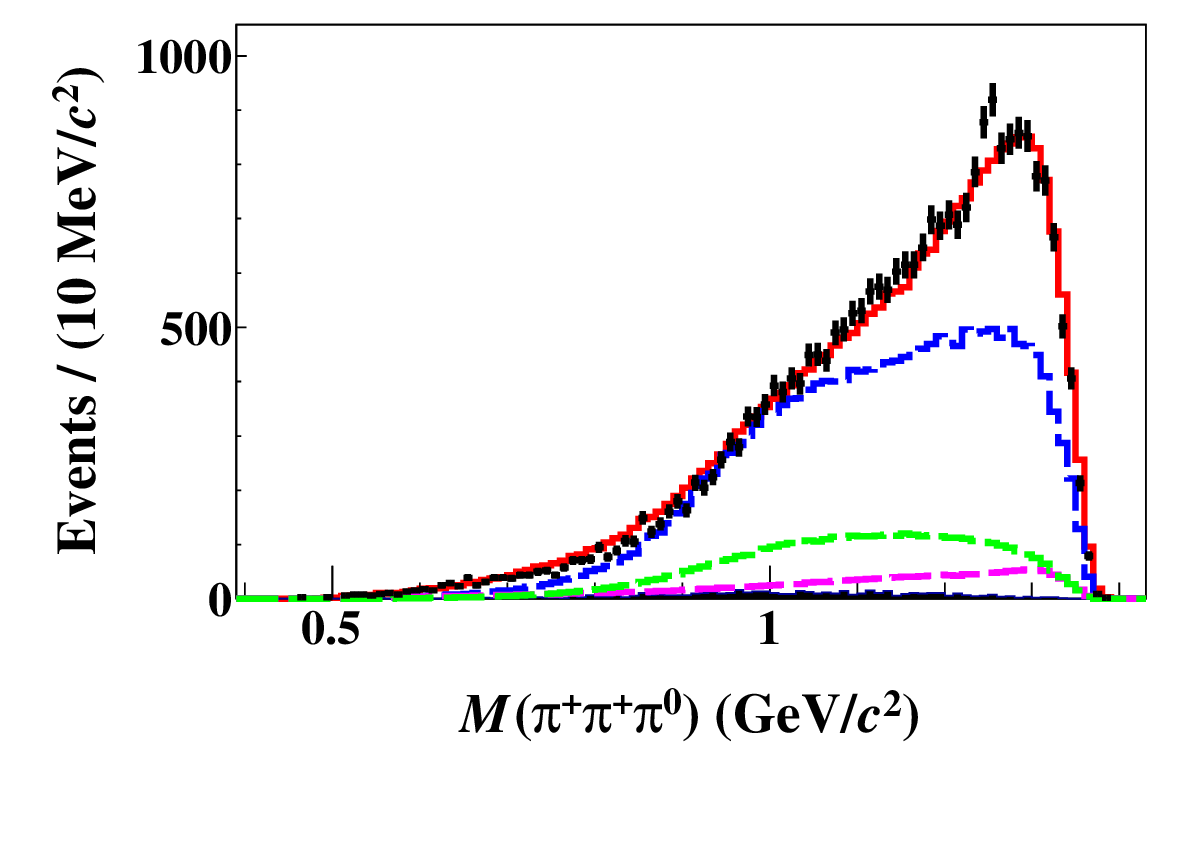}
    \includegraphics[width=0.45\textwidth]{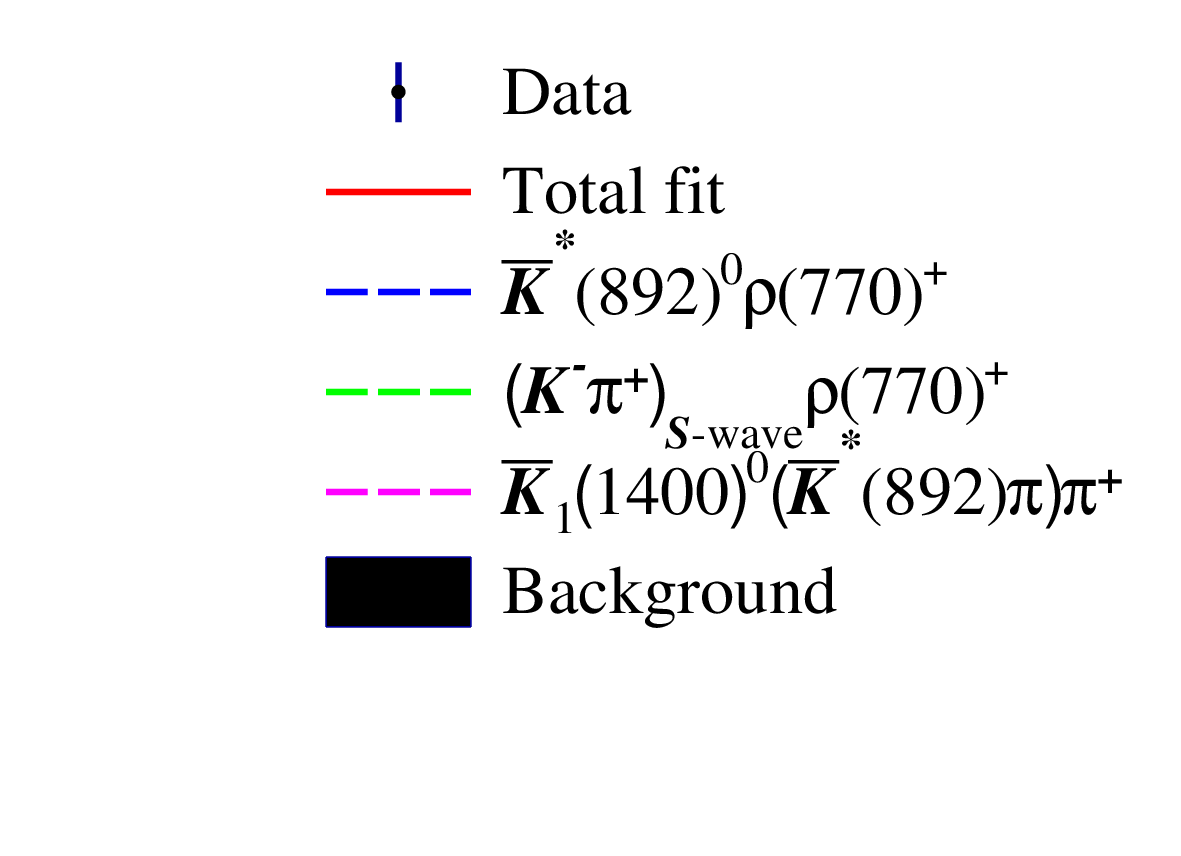}
	\caption{Projections of the baseline fit to the invariant-mass distributions. The combinations of two identical $\pi^+$ are added due to the exchange symmetry.}
    \label{fig:fitresult}
\end{figure}

In order to determine the statistical uncertainties of FFs, the amplitude coefficients are randomly sampled according to the covariant matrix. Then a Gaussian function is used to fit the distribution of each FF. The width of this function is assigned as the uncertainty of the corresponding FF. The phases, FFs, and statistical significances for different amplitudes are listed in
Table~\ref{fit-result}. The mass projections of the nominal fit are shown in Fig.~\ref{fig:fitresult}.

\begin{table}[htbp]\small
	\centering
	\begin{tabular}{|llc r@{ $\pm$ }c@{ $\pm$ }c c|}
	\hline
	 &Amplitude  &Phase~(rad) &\multicolumn{3}{c}{FFs~(\%)} &Significance~($\sigma$) \\
	\hline
		 &$D^{+}[S]\to \bar{K}^{*}(892)^0\rho(770)^+$  &0.0(fixed) &66.6 & 1.1 & 3.0 &$>10\sigma$\\
	       &$D^{+}[P]\to \bar{K}^{*}(892)^0\rho(770)^+$   &1.45 $\pm$ 0.04 $\pm$ 0.08 &1.9 & 0.2 &0.2 &$>10\sigma$ \\
		 &$D^{+}\to \bar{K}^{*}(892)^0\rho(770)^+$   &$-$ &68.5 & 1.1 & 2.6 &$>10\sigma$ \\
	\hline
	     &\makecell[l]{$D^{+}\to \bar{K}_1(1270)^0[S]\pi^+$, \\$\bar{K}_1(1270)^0\to K^-\rho(770)^+$}  &-0.09 $\pm$ 0.03 $\pm$ 0.03 &3.8 & 0.3 & 0.3 &$>10\sigma$ \\
	\hline
		 &$D^{+}\to \bar{K}_1(1400)^0[S]\pi^+$  &0.40 $\pm$ 0.02 $\pm$ 0.04 &7.5 &0.2 & 0.3 &$>10\sigma$ \\
			 &$D^{+}\to \bar{K}_1(1400)^0[D]\pi^+$  &-2.43$\pm$ 0.04 $\pm$ 0.04 &0.5 & 0.1 & 0.1 &$>10\sigma$ \\
		 &\makecell[l]{$D^{+}\to \bar{K}_1(1400)^0\pi^+$, \\$\bar{K}_1(1400)^0\to \bar{K}^{*}(892)\pi$} &$-$ &7.3 &0.2 & 0.3 &$>10\sigma$\\
	\hline
		 &\makecell[l]{$D^{+}\to \bar{K}(1460)^0\pi^+$, \\$\bar{K}(1460)^0\to \bar{K}^{*}(892)\pi$} &0.41 $\pm$ 0.04 $\pm$ 0.07 &5.1 & 0.2 & 0.3 &$>10\sigma$\\
	\hline
		 &\makecell[l]{$D^{+}\to \bar{K}^*(1680)^0\pi^+$, \\$\bar{K}^*(1680)^0\to \bar{K}^{*}(892)\pi$} &1.14 $\pm$ 0.04 $\pm$ 0.09 &3.9 & 0.4 & 0.8 &$>10\sigma$\\
	\hline
			 &$D^{+}\to (K^-\pi^+)_{S\rm{-wave}}\rho(770)^+$  &2.90 $\pm$ 0.02 $\pm$ 0.04 &18.3 &0.7 &0.7 &$>10\sigma$\\
	\hline
			 &\makecell[l]{$D^{+}\to \bar{K}(1460)^0\pi^+$, \\$\bar{K}(1460)^0\to K^-(\pi^+\pi^0)_V\pi$}   &-1.28 $\pm$ 0.08 $\pm$ 0.06 &8.4 &0.8 & 0.5 &$>10\sigma$\\
	\hline
			&\makecell[l]{$D^{+}\to \bar{K}(1460)^0\pi^+$, \\$\bar{K}(1460)^0\to (K^-\pi)_V\pi$}  &-2.31 $\pm$ 0.07 $\pm$ 0.06 &3.5 &0.5 & 0.3 &$>10\sigma$\\
	\hline
			&$D^{+}\to (K^-\rho(770)^+)_A\pi^+$  &-1.26 $\pm$ 0.04 $\pm$ 0.03 &1.8 & 0.1 & 0.1 &$>10\sigma$\\
	\hline
		         &$D^{+}\to (\bar{K}^{*}(892)\pi)_A\pi^+$  &-2.63 $\pm$ 0.05 $\pm$ 0.06 &0.8 & 0.1 & 0.1 &$>10\sigma$\\
	\hline
		         &$D^{+}\to (\bar{K}^{*}(892)^0\pi^+)_A\pi^0$  &-1.97 $\pm$ 0.05 $\pm$ 0.04 &0.8 & 0.2 & 0.4 &$>10\sigma$\\
	\hline
		    &$D^{+}\to (K^-(\pi^+\pi^0)_V)_P\pi^+$  &-1.13 $\pm$ 0.08 $\pm$ 0.13 &0.8 & 0.2 & 0.2 &$>10\sigma$\\
	\hline
		    &$D^{+}[S]\to (K^-\pi^+)_V\rho(770)^+$  &-1.88 $\pm$ 0.12 $\pm$ 0.11 &0.5 & 0.1 & 0.1 &9.3$\sigma$\\
	\hline	
    \end{tabular}
 \caption{Phases, FFs, and statistical significances for different amplitudes in  $D^+\to K^-\pi^+\pi^+\pi^0$. Groups of related amplitudes are separated by horizontal lines and the last row of each group gives the total fit fraction of the above components with interferences considered. The first and second uncertainties for the phases and FFs are statistical and systematic, respectively. 
The letters in bracket represent relative orbital angular momentum between resonances. The subscripts of $S$-wave denotes the $S$-wave that modeled by the LASS parameterization~\cite{KPsnew}, while the subscripts of $V$ $A$ and $P$ represent vector, axial-vector and pseudoscalar non-resonant components, respectively. 
The decay of $\bar{K}^*(892)$ includes both $K^{*}(892)^-$ and $\bar{K}^{*}(892)^0$, taking into account Clebsch-Gordan relations (refer to Appendix~\ref{CGrelation}).}

    \label{fit-result}
\end{table}

\subsection{Systematic uncertainties for the amplitude analysis}
\label{sec:PWA-Sys}
The systematic uncertainties for the amplitude analysis are described below and summarized in Table~\ref{tab:sys}. 
\begin{itemize}
\item[\uppercase\expandafter{\romannumeral1}] 
Amplitude model: \\
The masses and widths of resonances are adjusted by their corresponding uncertainties~\cite{PDG,K1270}. The GS lineshape of $\rho(770)^+$ is replaced with the RBW formula. The coupling constants of the $K\pi$ $S$-wave model are varied within their uncertainties given in ref.~\cite{KPsnew}. The changes of the phases and FFs are assigned as the associated systematic uncertainties. 

\item[\uppercase\expandafter{\romannumeral2}]
Effective radius: \\
The associated systematic uncertainties are estimated by repeating the fit procedure by varying the effective radii of the barrier, $R_r$, of the intermediate states and $D^+$ mesons by $R_r/\sqrt{12} \approx 1$ GeV$^{-1}$.
\item[\uppercase\expandafter{\romannumeral3}]  
Background: \\
        The background is determined from the inclusive MC sample, and the uncertainty from background is estimated by varying the $w_{\rm{bkg}}$ parameter in Eq.~(\ref{likelihood3}) within $\pm 1 \sigma$ of its statistical uncertainty.
\item[\uppercase\expandafter{\romannumeral4}] 
Simulation effects: \\
To estimate the uncertainties associated with $\gamma_{\epsilon}$, as defined in Eq.~(\ref{pwa:gamma}), the amplitude model is refitted by varying PID, tracking and $\pi^0$ reconstruction efficiencies according to their uncertainties. 
\item[\uppercase\expandafter{\romannumeral5}] 
Fit bias:\\
The uncertainty associated with the fit procedure is evaluated by studying signal MC samples. An ensemble of 600 signal MC samples are generated according to the results of the amplitude analysis to check the pull distribution. The pull variables, $\frac{V_{\rm fit}-V_{\rm input}}{\sigma_{\rm fit}}$,  are defined to evaluate the corresponding uncertainty, where $V_{\rm input}$ is the input value in the generator, and $V_{\rm fit}$ and $\sigma_{\rm fit}$ are the fit value and statistical uncertainty, respectively. The distribution of pull values for the 600 samples generated and fitted is expected to be a normal Gaussian distribution. Finally, the FFs and phases of all resonances as well as their statistical uncertainties are corrected by the fitted mean values of the pull distribution, and the uncertainty of the fitted mean values are assigned as the corresponding systematic uncertainties.
\item[\uppercase\expandafter{\romannumeral6}] 
Insignificant amplitudes:\\
The intermediate processes with statistical significances less than 5$\sigma$ are added one-by-one to the baseline fit and the largest variation of the phases and FFs is taken as the corresponding systematic uncertainty. Details are discussed in Appendix ~\ref{tested_amplitude}.

\end{itemize}

\begin{table}[htbp]
	\centering
	\begin{tabular}{|lcccccccc|}
	\hline
		\   &\multicolumn{8}{c|}{Source}    \\
	\hline
		Amplitude                                             &\    &I    &II    &III   &IV   &V  &VI &Total\\
			\hline

		$D^{+}[S]\to \bar{K}^{*}(892)^0\rho(770)^+$ 
		&FF &0.50 	&0.31 &0.08 &0.05 &0.06 &2.65 &2.72\\ 
		\multirow{2}{*}{$D^{+}[P]\to \bar{K}^{*}(892)^0\rho(770)^+$} 
		&$\phi$ &1.70 	&0.34 &0.11 &0.08 &0.06 &0.95 &1.98\\ 
		&FF &0.84 	&0.10 &0.01 &0.03 &0.05 &0.20 &0.87\\
		$D^{+}\to \bar{K}^{*}(892)^0\rho(770)^+$ 
		&FF &0.48 	&0.26 &0.07 &0.04 &0.04 &2.33 &2.40\\
	\hline
		\multirow{2}{*}{\makecell[l]{$D^{+}\to \bar{K}_1(1270)^0\pi^+$, \\$\bar{K}_1(1270)^0[S]\to K^- \rho(770)^+$}} 
		&$\phi$ &0.90 	&0.22 &0.04 &0.09 &0.05 &0.22 &0.96\\ 
		&FF &0.31 	&0.15 &0.10 &0.15 &0.06 &0.95 &1.03\\
	\hline
		\multirow{2}{*}{\makecell[l]{$D^{+}\to \bar{K}_1(1400)^0\pi^+$ , \\$\bar{K}_1(1400)^0[S]\to \bar{K}^{*}(892)\pi$}} 
		&$\phi$ &1.80 	&0.33 &0.03 &0.01 &0.06 &0.70 &1.96\\ 
		&FF &1.02 	&0.15 &0.01 &0.01 &0.06 &0.85 &1.34\\
		\multirow{2}{*}{\makecell[l]{$D^{+}\to \bar{K}_1(1400)^0\pi^+$ , \\$\bar{K}_1(1400)^0[D]\to \bar{K}^{*}(892)\pi$}} 
		&$\phi$ &1.06 &0.24 &0.07 &0.03 &0.05 &0.07 &1.09\\ 
		&FF &0.20 	&0.10 &0.10 &0.01 &0.05 &0.01 &0.25\\
		$D^{+}\to \bar{K}_1(1400)^0\pi^+$ 
		&FF &0.96 	&0.20 &0.05  &0.05 &0.06 &0.80 &1.27\\
	\hline
		\multirow{2}{*}{\makecell[l]{$D^{+}\to \bar{K}(1460)^0\pi^+$, \\$\bar{K}(1460)^0\to \bar{K}^{*}(892)\pi$}} 
		&$\phi$ &1.55 	&0.05 &0.17 &0.03 &0.05 &0.66 &1.70\\ 
		&FF &1.49 	&0.30 &0.15 &0.01 &0.05 &0.35 &1.57\\
		\multirow{2}{*}{\makecell[l]{$D^{+}\to \bar{K}^{*}(1680)^{0}\pi^+$, \\$\bar{K}^{*}(1680)^{0}\to \bar{K}^{*}(892)\pi$}} 
		&$\phi$ &2.19 &0.21 &0.02 &0.01 &0.06 &0.14 &2.20\\ 
		&FF &1.89 	&0.43 &0.03 &0.11 &0.06 &0.05 &1.94\\
		\multirow{2}{*}{$D^{+}\to (K^-\pi^+)_{S\rm{-wave}}\rho(770)^+$} & 
		$\phi$ &1.76 &0.52 &0.06 &0.04 &0.06 &0.38 &1.88\\ 
		&FF &0.95 &0.15 &0.03 &0.08 &0.06 &0.15 &0.98\\
	\hline
		\multirow{2}{*}{\makecell[l]{$D^{+}\to \bar{K}(1460)^0\pi^+$, \\$\bar{K}(1460)^0\to K^-(\pi^+\pi^0)_V$}} 
		&$\phi$ &0.42 &0.26 &0.01 &0.05 &0.06 &0.52 &0.72\\ 
		&FF &0.51 &0.03 &0.02 &0.14 &0.06 &0.32 &0.62\\
	\hline
		\multirow{2}{*}{\makecell[l]{$D^{+}\to \bar{K}(1460)^0\pi^+$, \\$\bar{K}(1460)^0\to (K^-\pi)_{V}\pi$}} 
		&$\phi$ &0.76 &0.37 &0.03 &0.06 &0.06 &0.14 &0.86\\ 
		&FF &0.34 	&0.60 &0.02 &0.04 &0.06 &0.56&0.89\\
	\hline
		\multirow{2}{*}{$D^{+}\to (K^-\rho(770)^+)_A\pi^+$} & 
		$\phi$ &0.43 &0.53 &0.04 &0.05 &0.04 &0.31 &0.75\\ 
		&FF &0.47 &0.70 &0.01 &0.20 &0.05 &0.40 &0.96\\
	\hline
		\multirow{2}{*}{$D^{+}\to (\bar{K}^{*}(892)\pi)_A\pi^+$} & 
		$\phi$ &1.20 &0.05 &0.08 &0.10 &0.06 &0.24 &1.23\\ 
		&FF &0.72 &0.20 &0.01 &0.01 &0.06 &0.30 &0.80\\
		\multirow{2}{*}{$D^{+}\to (\bar{K}^{*}(892)^0\pi^+)_A\pi^0$} & 
		$\phi$ &0.78 &0.15 &0.05 &0.10 &0.06 &0.29 &0.86\\ 
		&FF &0.96 &0.30 &0.10 &0.10 &0.05 &1.95 &2.20\\
    \hline
		\multirow{2}{*}{$D^{+}\to (K^-(\pi^+\pi^-)_V)_P\pi^+$} & 
		$\phi$ &1.39 &0.75 &0.04 &0.12 &0.06 &0.26 &1.61\\ 
		&FF &0.74 &0.20 &0.01 &0.10 &0.06 &0.50 &0.92\\
    \hline
		\multirow{2}{*}{$D^{+}[S]\to (K^-\pi^+)_{V}\rho(770)^+$} & 
		$\phi$ &0.69 &0.42 &0.08 &0.01 &0.06 &0.43 &0.92\\ 
		&FF &0.47 &0.20 &0.10 &0.01 &0.05 &0.60 &0.80\\

    \hline
	\end{tabular}
		\caption{Systematic uncertainties on the phases and FFs for the different components in the amplitude model in units of the corresponding statistical uncertainties.  (I) Amplitude model, (II) Effective radius, (III) Background, (IV) Simulation effects, (V) Fit bias, and (VI) Insignificant amplitudes.}
	\label{tab:sys}

\end{table}
\section{Measurement of the branching fraction}
The BF of  $D^+ \to K^-\pi^+\pi^+\pi^0$ is measured with the DT technique using the same tag modes and event selection criteria as those described in Sec.~\ref{ST-selection}.

For a given ST mode, we have
\begin{equation}
  N_{\text{tag}}^{\text{ST}} = 2N_{D^{+}D^{-}}\cdot \mathcal{B}_{\text{tag}}\cdot\epsilon_{\text{tag}}^{\text{ST}}\,, \label{eq-ST}
\end{equation}
\begin{equation}
  \begin{array}{lr}
    N_{\text{tag,sig}}^{\text{DT}}=2N_{D^{+}D^{-}}\cdot \mathcal{B}_{\pi^0\to \gamma \gamma}\cdot \mathcal{B}_{\text{tag}}\cdot\mathcal{B}_{\text{sig}}\cdot \epsilon_{\text{tag,sig}}^{\text{DT}}\,,
  \end{array}
  \label{eq-DT}
\end{equation}
where $N_{\text{tag}}^{\text{ST}}$ is the ST yield for a specific tag mode, $N_{D^{+}D^{-}}$ is the total number of $D^{+}D^{-}$ pairs produced from $e^{+}e^{-}$ collisions, $\mathcal{B}_{\text{tag}}$ and $\epsilon_{\text{tag}}^{\text{ST}}$ are the BF and the ST efficiency for tag mode, $N_{\text{tag,sig}}^{\text{DT}}$ is the DT yield, $\mathcal{B}_{\text{sig}}$ and $\epsilon_{\text{tag,sig}}^{\text{DT}}$ are the BF of the signal mode and the efficiency for simultaneously reconstructing the signal and tag modes, $\mathcal{B}_{\pi^0\to \gamma \gamma}$ is the BF of $\pi^0 \to \gamma \gamma$ obtained from PDG~\cite{PDG}. Combining the two equations above, the absolute BF of  $D^+ \to K^-\pi^+\pi^+\pi^0$ is 
\begin{equation}
  \mathcal{B}_{\text{sig}} = \frac{N_{\text{tag,sig}}^{\text{DT}}}{\begin{matrix} \mathcal{B}_{\pi^0\to \gamma \gamma}\cdot N_{\text{tag}}^{\text{ST}}\cdot \epsilon^{\text{DT}}_{\text{tag,sig}}/\epsilon_{\text{tag}}^{\text{ST}}\end{matrix}}\,.\label{BR-formula}
\end{equation} 

\begin{figure*}[!htbp]
  \centering
   \includegraphics[width=0.5\textwidth]{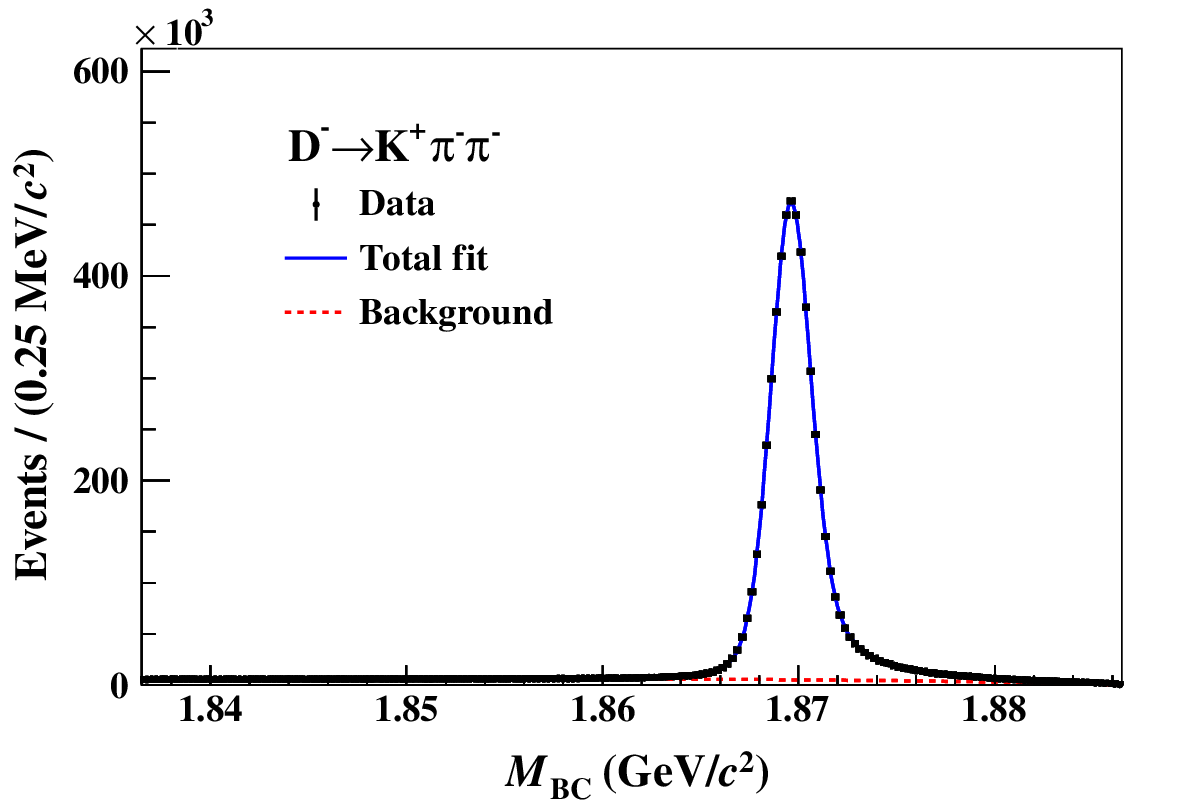} 
 \caption{Fit to the $M_{\rm BC}$ distribution of the ST candidates. }
  \label{STFit}
\end{figure*}

The value of $N_{\text{tag}}^{\text{ST}}$ is obtained from a one-dimensional binned fit to the $M_{\rm BC}$ distribution, as shown in Fig.~\ref{STFit}. The signal shape is modeled by the MC-simulated shape convolved with a double-Gaussian function describing the resolution difference between data and MC simulation, and the background shape is described by the ARGUS function~\cite{ARGUS}. The corresponding $\epsilon_{\text{tag}}^{\text{ST}}$ is estimated with the inclusive MC sample.

\begin{figure}[htbp]
  \centering
  \includegraphics[width=0.5\textwidth]{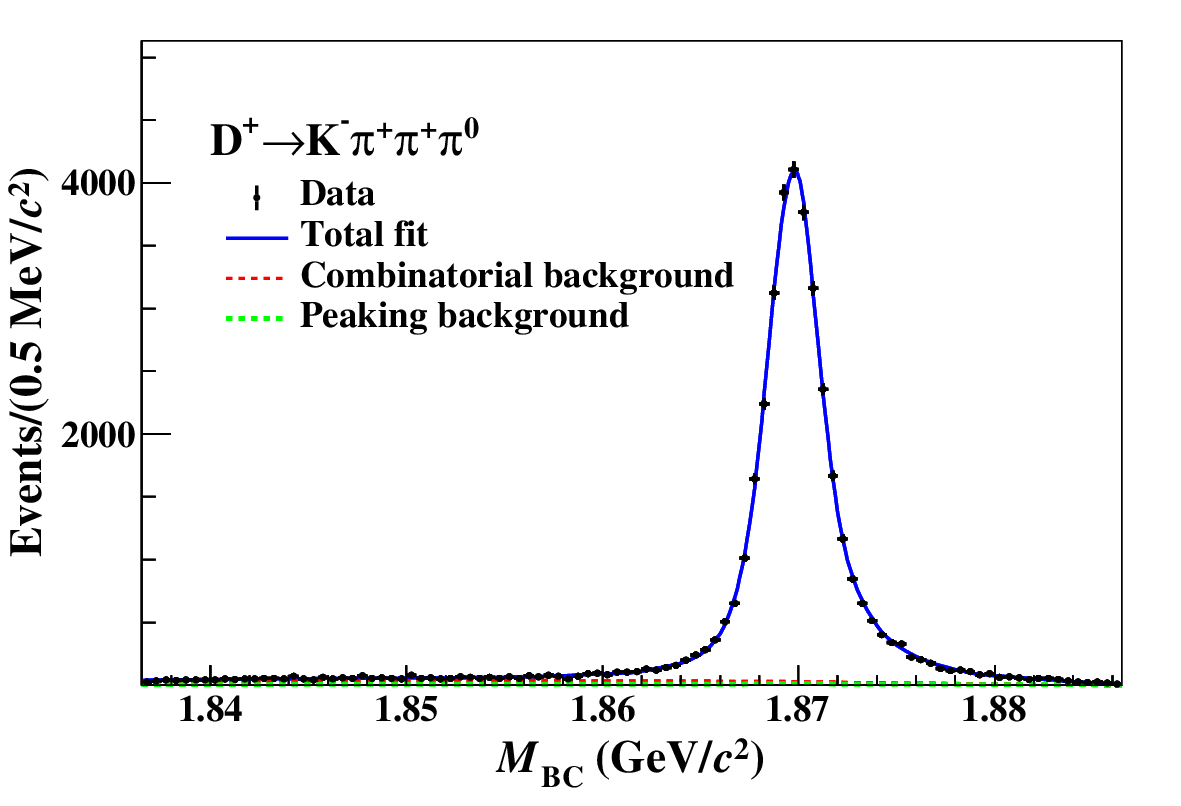}
\caption{Fit to the $M_{\rm BC}$ distribution of the DT candidates.}
  \label{DTFit}
\end{figure}

The DT yield is determined to be  $N_{\text{tag,sig}}^{\text{DT}}=35481 \pm 220$ through a fit to the $M_{\rm BC}$ distribution, as depicted in Fig.~\ref{DTFit}. Here, the signal shape is modeled by the MC-simulated shape convolved with a double-Gaussian function, and the background shape is described by the ARGUS function~\cite{ARGUS}. There are some peaking backgrounds from the processes $D^+ \to K \pi e^+ \nu_e$ and $D^+ \to K \pi \mu^+ \nu_{\mu}$ , and they are described by the MC-simulated shape, and the corresponding contributions are
fixed to the estimation from the MC simulation in the fit. $\epsilon^{\text{DT}}_{\text{tag,sig}}$ is determined with the signal MC sample in which the $D^+ \to K^-\pi^+\pi^+\pi^0$ events are generated according to the result of the amplitude analysis.
The values of these parameters are summarized in Table~\ref{ST-eff}. 

\begin{table}[htbp]
  \label{ST-eff}
    \begin{center}
    \resizebox{\textwidth}{!}{
      \begin{tabular}{lccccc}
        \toprule\toprule
        Tag mode          & $\Delta E$~(MeV)  &$N^{\text{ST}}_{\text{tag}}$  & $\epsilon^{\text{ST}}_{\text{tag}}(\%)$  &$N_{\text{tag,sig}}^{\text{DT}}$ &$\epsilon^{\text{DT}}_{\text{tag,sig}}(\%)$ \\
        \hline
	$D^{-} \rightarrow K^{+}\pi^{-}\pi^{-}$               &$[-25,24]$  &$2215326 \pm 1589$  &$ 52.44 \pm 0.01$  &$35481 \pm 220$ &$14.03 \pm 0.01$         \\  
        \bottomrule\bottomrule
      \end{tabular}
      }
    \end{center}
      \caption{The energy difference requirements, ST yields ($N^{\text{ST}}_{\text{tag}}$), ST efficiency ($\epsilon^{\text{ST}}_{\text{tag}}$), DT yields~($N_{\text{tag,sig}}^{\text{DT}}$) and DT efficiency ($\epsilon^{\text{DT}}_{\text{tag,sig}}$). The uncertainties are statistical only.}

\end{table}

The systematic uncertainties for the branching fraction measurement are described below and  summarized in Table~\ref{BF-Sys}.


\begin{itemize}
\item ST $D^{-}$ candidates:

The uncertainty in the yield of ST $D$ mesons is assigned to be 0.1\% from studies that involve varying the signal shape, background shape, and  floating the parameters of the Gaussian in the fit~\cite{arXiv2312}.

\item Tracking and PID:

The tracking and PID efficiencies of $\pi^{\pm}$ and $K^{\pm}$ are investigated with DT hadronic $D\bar{D}$ events of the decays $D^0\to K^-\pi^+$, $K^-\pi^+\pi^0$, $K^-\pi^+\pi^+\pi^-$ versus $\bar{D}^0\to K^+\pi^-$, $K^+\pi^-\pi^0$, $K^+\pi^-\pi^-\pi^+$, and $D^+\to K^-\pi^+\pi^+$ versus $D^- \to K^+\pi^-\pi^-$.
The data-MC efficiency ratios for $\pi$ and $K$ tracking~(PID) are found to be 0.996 $\pm$ 0.002 and 0.994 $\pm$ 0.003~(0.998 $\pm$ 0.001 and 0.968 $\pm$ 0.003). 
After correcting the MC efficiencies to data by these factors, the statistical uncertainties of the correction parameters are assigned as the systematic uncertainties. These are 0.4\% and 0.3\%~(0.2\%  and 0.3\%) for the $\pi$ and $K$ tracking~(PID), respectively. 

\item $\pi^{0}$ reconstruction:

The data-MC efficiency ratio for $\pi^0$ reconstruction is $0.997 \pm 0.004$, which is measured with samples of $D^0 \to K^-\pi^+, K^-\pi^+\pi^+\pi^-$ versus $\bar{D}^0 \to K^+\pi^-\pi^0, K_S^0\pi^0$ hadronic decays. After correcting the efficiency of $\pi^{0}$ reconstruction by this factor, the associated systematic uncertainty is assigned as 0.4\%.

\item MC sample size:

The statistical uncertainty arising from the limited size of the MC sample is 0.1\%.

\item Amplitude  model:

The uncertainty associated with the amplitude model is estimated by varying the fitted parameters based on the covariance matrix. 
The masses and widths of the intermediate resonances and $R_r$ values are randomized according to a Gaussian distribution. The distribution of 300 efficiencies arising from this procedure is fitted by a Gaussian and the deviation from the nominal mean value, 0.8\%, is taken as the systematic uncertainty.

\item The assumed BF:

The BF of $\pi^0 \to \gamma \gamma$ is well known~\cite{PDG}, and the uncertainty on this quantity induces a negligible uncertainty in the analysis.   

\item Fit procedure:

The signal MC accurately describes the distribution in data,  and the fit model includes a convolution of a free Gaussian function to account for residual  discrepancies in data-MC resolution. 
Therefore, the primary source of systematic uncertainty in the fit procedure stems from the knowledge of the background shape. The uncertainty is estimated by shifting the end-point of the ARGUS function by $\pm 0.2$~MeV, and varying the fixed value for the peaking background within a range of $\pm 1 \sigma$ of its statistical uncertainty. The largest deviation in the measured BF, which is 0.1\%, is assigned as the corresponding systematic uncertainty.

\end{itemize}

After correcting for the differences in $\pi^+$ tracking, PID and $\pi^0$ reconstruction efficiencies between data and MC simulation, the BF of $D^+ \to K^-\pi^+\pi^+\pi^0$ is determined to be $\mathcal{B}(D^+ \to K^-\pi^+\pi^+\pi^0)$ = \BF. 

\begin{table}[htbp]
  \label{BF-Sys}
  \begin{center}
    \begin{tabular}{cccc}
      \toprule\toprule
      Source   &Uncertainty (\%)\\
      \hline
      ST $D^{-}$ candidates               & 0.1 \\
      Tracking      & 0.7 \\
      PID           & 0.5\\
      $\pi^0$ reconstruction              &0.4 \\
      MC sample size                       & 0.1\\
      Amplitude  model                        & 0.8 \\
      $\mathcal{B}(\pi^0 \to \gamma \gamma)$		&Negligible\\
      Fit procedure		&0.1~\\
      \hline
      Total                               & 1.2 \\
      \bottomrule\bottomrule
    \end{tabular}
  \end{center}
  \caption{Relative systematic uncertainties in the BF measurement.}
\end{table}

\section{Summary}
An amplitude analysis of the Cabibbo-favoured decay $D^{+} \to K^-\pi^{+}\pi^{+}\pi^{0}$ has been performed  using 7.93 $\rm{fb}^{-1}$ of $e^+e^-$ collision data collected with the BESIII detector at the center-of-mass energy of 3.773 GeV. With a detection efficiency based on the results of the amplitude analysis, we obtain $\mathcal{B}(D^+\to K^-\pi^+\pi^+\pi^0)$ = \BF. The result is consistent with the  value of $\mathcal{B}(D^+\to K^-\pi^+\pi^+\pi^0) = (5.98\pm0.08_{\rm{stat.}}\pm0.16_{\rm{syst.}})\%$ measured by the CLEO collaboration~\cite{bf}.
Combining the FFs listed in Table~\ref{fit-result}, the BFs for the intermediate processes are calculated with $\mathcal{B}_i = FF_i \times \mathcal{B}(D^+\to K^-\pi^+\pi^+\pi^0)$, and the obtained results are listed in Table~\ref{tab:bfamp}. 

According to the amplitude analysis, the dominant intermediate process is $D^+ \to \bar{K}^{*}(892)^0\rho(770)^+\to K^-\pi^+\pi^+\pi^0$, with a  BF of $(4.15\pm0.07_{\rm{stat.}}\pm0.17_{\rm{syst.}})\%$. After applying the isospin symmetry assumption to the decays of $\bar{K}^{*}(892)^0\to K^-\pi^+$ and $\bar{K}^{*}(892)^0\to \bar{K}^0\pi^0$, the absolute BF of $D^+ \to \bar{K}^{*}(892)^0\rho(770)^+$ is determined to be $(6.23\pm0.11_{\rm{stat.}}\pm0.25_{\rm{syst.}})\%$.   As can be seen from Table~\ref{tab:bfcmp}, 
this result is consistent with previous measurements from MARK-III~\cite{MK} and BESIII~\cite{JHEP09077}, but is a factor of 10.1 and 2.1 times more precise, respectively.

The measured BF of $D^+ \to \bar{K}_1(1400)^0\pi^+$ is consistent with the previous BESIII result~\cite{JHEP09077} within 1.5$\sigma$, but the precision is improved by a factor of 4.4. Information about the two $K_1$ states in this decay also provides inputs to further investigations of the mixing of the axial-vector kaon mesons~\cite{CHY:2012ggx,PRD125}.  

\begin{table}[t]
\setlength{\abovecaptionskip}{0.cm}
\setlength{\belowcaptionskip}{-0.2cm}
  \begin{center}
    \begin{tabular}{|lc|}
      \hline
      Intermediate process & BF ($10^{-2}$)\\
      \hline
      $D^{+}\to \bar{K}^{*}(892)^0\rho(770)^+, \bar{K}^{*}(892)^0\to K^-\pi^+, \rho(770)^+\to \pi^{+}\pi^{0}$ &4.15 $\pm$ 0.07 $\pm$ 0.17 \\        
       \hline
      $D^{+}\to \bar{K}_1(1270)^0\pi^+,  \bar{K}_1(1270)^0\to K^-\rho(770)^+, \rho(770)^+\to \pi^{+}\pi^{0}$ &0.23 $\pm$ 0.02 $\pm$ 0.02 \\
 \hline
       $D^{+}\to \bar{K}_1(1400)^0\pi^+, \bar{K}_1(1400)^0\to \bar{K}^{*}(892)\pi, \bar{K}^{*}(892)\to K\pi$ &0.44 $\pm$ 0.01 $\pm$ 0.02 \\
 \hline
      $D^{+}\to \bar{K}(1460)^0\pi^+, \bar{K}(1460)^0\to \bar{K}^{*}(892)\pi,\bar{K}^{*}(892)\to K\pi$ &0.31 $\pm$ 0.01 $\pm$ 0.02 \\
 \hline
       $D^{+}\to \bar{K}(1680)^{*0}\pi^+, \bar{K}(1680)^{*0}\to \bar{K}^{*}(892)\pi, \bar{K}^{*}(892)\to K\pi$ &0.23 $\pm$ 0.02 $\pm$ 0.02 \\
       \hline
      $D^{+}\to (K^-\pi^+)_{S\rm{-wave}}\rho(770)^+, \rho(770)^+\to \pi^{+}\pi^{0}$ &1.11 $\pm$ 0.04 $\pm$ 0.04\\
\hline
      $D^{+}\to \bar{K}(1460)^0\pi^+, \bar{K}(1460)^0 \to K^-(\pi^+\pi^0)_V$ &0.51 $\pm$ 0.04 $\pm$0.03\\
 \hline
      $D^{+}\to \bar{K}(1460)^0\pi^+, \bar{K}(1460)^0 \to (K^-\pi)_{V}\pi$ &0.22 $\pm$ 0.03 $\pm$ 0.03\\
 \hline      
      $D^{+}\to (K^-\rho(770)^+)_A\pi^+, \rho(770)^+\to \pi^{+}\pi^{0}$ &0.11 $\pm$ 0.01 $\pm$ 0.01 \\
 \hline      
      $D^{+}\to (\bar{K}^{*}(892)\pi)_A\pi^+, \bar{K}^{*}(892)\to K\pi$ &0.05 $\pm$ 0.01 $\pm$ 0.01 \\
 \hline
      $D^{+}\to (\bar{K}^{*}(892)^0\pi^+)_A\pi^0, \bar{K}^{*}(892)^0\to K^-\pi^+$ &0.05 $\pm$ 0.01 $\pm$0.02\\
 \hline 
      $D^{+}\to (K^-\pi^+)_V\rho(770)^+, \rho(770)^+\to \pi^{+}\pi^{0}$ &0.03 $\pm$ 0.01 $\pm$0.01\\      
      \hline
      $D^{+}\to (K^-(\pi^+\pi^0)_V)_P\pi^+$ &0.05 $\pm$ 0.01 $\pm$0.01\\
 \hline           
    \end{tabular}
  \end{center}
      \caption{The BFs of various intermediate processes in $D^+\to K^-\pi^{+}\pi^{+}\pi^{0}$. The first and second uncertainties are statistical and systematic, respectively. }
\label{tab:bfamp}
\end{table}

\begin{table}[t]
\setlength{\abovecaptionskip}{0.cm}
\setlength{\belowcaptionskip}{-0.2cm}
  \begin{center}
    \begin{tabular}{|lc|}
      \hline
      Decay channel and Collaboration &$\mathcal{B}(D^{+}\to \bar{K}^{*}(892)^0\rho(770)^+)$ ($\times 10^{-2}$) \\
      \hline
      $D^+\to K^-\pi^{+}\pi^{+}\pi^{0}$, current analysis &6.23 $\pm$ 0.11 $\pm$ 0.25\\         \hline
      $D^+\to K^-\pi^{+}\pi^{+}\pi^{0}$,  MARK-III~\cite{MK} &7.2 $\pm$ 1.8 $\pm$ 2.1\\         \hline
      $D^+\to K_S^0\pi^{+}\pi^{0}\pi^{0}$,  BESIII~\cite{JHEP09077} &5.82 $\pm$ 0.49 $\pm$ 0.29\\         \hline
    \end{tabular}
  \end{center}
      \caption{Comparison of the BFs of the intermediate processes $D^{+}\to \bar{K}^{*}(892)^0\rho(770)^+$ in the $D$ hadronic decay. The first and second uncertainties are statistical and systematic, respectively.}

\label{tab:bfcmp}
\end{table}

\acknowledgments

The BESIII Collaboration thanks the staff of BEPCII and the IHEP computing center for their strong support. This work is supported in part by National Key R\&D Program of China under Contracts Nos. 2020YFA0406300, 2020YFA0406400, 2023YFA1606000; National Natural Science Foundation of China (NSFC) under Contracts Nos. 11635010, 11735014, 11935015, 11935016, 11935018, 12025502, 12035009, 12035013, 12061131003, 12192260, 12192261, 12192262, 12192263, 12192264, 12192265, 12221005, 12225509, 12235017, 12361141819; the Chinese Academy of Sciences (CAS) Large-Scale Scientific Facility Program; the CAS Center for Excellence in Particle Physics (CCEPP); Joint Large-Scale Scientific Facility Funds of the NSFC and CAS under Contract Nos. U2032108, U1832207; Shanghai Leading Talent Program of Eastern Talent Plan under Contract No. JLH5913002; 100 Talents Program of CAS; The Institute of Nuclear and Particle Physics (INPAC) and Shanghai Key Laboratory for Particle Physics and Cosmology; German Research Foundation DFG under Contracts Nos. FOR5327, GRK 2149; Istituto Nazionale di Fisica Nucleare, Italy; Knut and Alice Wallenberg Foundation under Contracts Nos. 2021.0174, 2021.0299; Ministry of Development of Turkey under Contract No. DPT2006K-120470; National Research Foundation of Korea under Contract No. NRF-2022R1A2C1092335; National Science and Technology fund of Mongolia; National Science Research and Innovation Fund (NSRF) via the Program Management Unit for Human Resources \& Institutional Development, Research and Innovation of Thailand under Contracts Nos. B16F640076, B50G670107; Polish National Science Centre under Contract No. 2019/35/O/ST2/02907; Swedish Research Council under Contract No. 2019.04595; The Swedish Foundation for International Cooperation in Research and Higher Education under Contract No. CH2018-7756; U. S. Department of Energy under Contract No. DE-FG02-05ER41374.

\bibliographystyle{JHEP}
\bibliography{references}

\clearpage
\appendix
\section{$M_{\rm{BC}}^{\rm{sig}}$ versus $M_{\rm{BC}}^{\rm{tag}}$ two-dimensional fit}
\label{2dfit}
The signal yields of DT candidates are determined by a two-dimensional (2D) maximum likelihood bin fit to the distribution of $M_{\rm{BC}}^{\rm{sig}}$ versus $M_{\rm{BC}}^{\rm{tag}}$. Signal events with both and signal sides reconstructed correctly concentrate around $M_{\rm{BC}}^{\rm{sig}} = M_{\rm{BC}}^{\rm{tag}} = M_D$, where $M_D$ is the known $D$ mass~\cite{PDG}.  We define three kinds of background. Candidates with correctly reconstructed $D^+$~(or $D^-$) and incorrectly reconstructed $D^-$~(or $D^+$) are BKGI, which appear around the lines $M_{\rm{BC}}^{\rm{sig}}$ or $M_{\rm{BC}}^{\rm{tag}}$ = $M_D$. Other candidates appearing around the diagonal are mainly from the $D^0\bar{D}^0$ mispartition and the $e^+e^- \to q\bar{q}$ processes (BKGII). The remaining combinatorial backgrounds mainly come from candidates reconstructed incorrectly on both sides (BKGIII). The PDFs for the different components used in the fit are given below:
\begin{itemize}
\item \textbf{Signal: s($x, y$),}
\item \textbf{BKGI: $b_1(x) \cdot {\mathcal A}$rgus($y; m_0, c, p) + b_2(y) \cdot$ ${\mathcal A}$rgus($x; m_0, c, p$),}
\item \textbf{BKGII:  ${\mathcal A}$gus($(x+y)/\sqrt{2}; m_0, c, p) \cdot g((x-y)/\sqrt{2})$,}
\item \textbf{BKGIII: ${\mathcal A}$rgus($x; m_0, c, p$) $\cdot$ ${\mathcal A}$rgus($y; m_0, c, p$).}
 \end{itemize}

The signal shape s($x, y$) is described by the 2D MC-simulated shape convolved with a 2D Gaussian. The parameters of the Gaussian function are obtained by one-dimensional~(1D) fit on $M_{\rm{BC}}$ in signal and tag sides respectively, and are fixed in 2D fit. For BKGI, $b_{1,2}(x, y)$ is described by the 1D MC-simulated shape convolved with a Gaussian function, ${\mathcal A}$rgus($x, y$) is the ARGUS function~\cite{ARGUS}. For BKGII, it is described by an ARGUS function in the diagonal axis multiplied by a Gaussian function in the anti-diagonal axis. For BKGIII, it constructed by an ARGUS function~\cite{ARGUS} in $M_{\rm{BC}}^{\rm{sig}}$ multiplied by an ARGUS function in $M_{\rm{BC}}^{\rm{tag}}$. In the fit, the parameters $m_0$ and $p$ for the ARGUS function~\cite{ARGUS} is fixed at 1.8865 GeV/$c^2$ and 0.5, respectively.

\section{Clebsch-Gordan relations}
\label{CGrelation}
Considering the isospin relationship in hadron decays, some amplitudes are fixed by CG relations, as listed in table~\ref{tab:CG}. The amplitudes with fixed relations share the same magnitude $(\rho)$ and phase $(\phi)$. 

\begin{table}[h]
  \begin{center}
    \begin{tabular}{|clc|}
      \hline
      Index                &Amplitude              &Relation\\
      \hline
      $A_1$   &$D^{+}\to \bar{K}_1(1400)^0\pi^+$, $\bar{K}_1(1400)^0\to \bar{K}^{*}(892)^0\pi^0, \bar{K}^{*}(892)^0 \to K^-\pi^+$            & \\
      $A_2$   &$D^{+}\to \bar{K}_1(1400)^0\pi^+$, $\bar{K}_1(1400)^0\to K^{*}(892)^-\pi^+, K^{*}(892)^- \to K^-\pi^0$                & \\
      $A$    &$D^{+}\to \bar{K}_1(1400)^0\pi^+$, $\bar{K}_1(1400)^0\to K^{*}(892) \pi,K^{*}(892) \to K\pi$                &$A_1-A_2$ \\
			\hline
	  $A_1$   &$D^{+}\to \bar{K}(1460)^0\pi^+$, $\bar{K}(1460)^0\to \bar{K}^{*}(892)^0\pi^0, \bar{K}^{*}(892)^0 \to K^-\pi^+$            & \\
      $A_2$   &$D^{+}\to \bar{K}(1460)^0\pi^+$, $\bar{K}(1460)^0\to K^{*}(892)^-\pi^+, K^{*}(892)^- \to K^-\pi^0$                & \\
      $A$    &$D^{+}\to \bar{K}(1460)^0\pi^+$, $\bar{K}(1460)^0\to K^{*}(892) \pi, K^{*}(892) \to K\pi$                &$A_1-A_2$ \\
			\hline
	  $A_1$   &$D^{+}\to \bar{K}^{*}(1680)^{0}\pi^+$, $\bar{K}^{*}(1680)^{0}\to \bar{K}^{*}(892)^0\pi^0, \bar{K}^{*}(892)^0 \to K^-\pi^+$            & \\
      $A_2$   &$D^{+}\to \bar{K}^{*}(1680)^{0}\pi^+$, $\bar{K}^{*}(1680)^{0}\to K^{*}(892)^-\pi^+, K^{*}(892)^- \to K^-\pi^0$                & \\
      $A$    &$D^{+}\to \bar{K}^{*}(1680)^{0}\pi^+$, $\bar{K}^{*}(1680)^{0}\to K^{*}(892) \pi, K^{*}(892) \to K\pi$                &$A_1-A_2$ \\
      \hline
    \end{tabular}
  \end{center}
        \caption{The CG relations assumed in the analysis.}
          \label{tab:CG}
\end{table}

\renewcommand\thesection{\Alph{section}}
\section{Other intermediate processes tested}
\label{tested_amplitude}
In this section, we list the significance and interference of the other possible combinations of the extra intermediate resonances that are considered in the amplitude analysis. Note that, the parameterized $\bar{K}^{*}(700)^0$ with T-matrix Pole~(POLE) function as the propagator is labeled as $\bar{K}^{*}(700)^0(\rm{POLE})$, and the parameterized $\bar{K}^{*}(700)^0$ with relativistic Breit-Wigner~(RBW) function as the propagator is labeled as $\bar{K}^{*}(700)^0(\rm{RBW})$.\\

\begin{itemize}
\setlength{\itemsep}{0pt}
\item \textbf{Cascade amplitudes}
\item[-]      $D^{+}[D]\to \bar{K}^{*}(892)^0\rho(770)^+$~(2.6$\sigma)$,
\item[-]      $D^{+}[S]\to \bar{K}^{*}(892)^0\rho(1450)^+$~(4.4$\sigma)$,
\item[-]      $D^{+}[P]\to \bar{K}^{*}(892)^0\rho(1450)^+$~(3.8$\sigma)$,
\item[-]      $D^{+}[D]\to \bar{K}^{*}(892)^0\rho(1450)^+$~(3.5$\sigma)$,
\item[-]      $D^{+}\to \bar{K}^{*}(700)^0(\rm{RBW})\rho(770)^+$(3.0$\sigma)$,
\item[-]      $D^{+}\to \bar{K}^{*}(700)^0(\rm{RBW})\rho(1450)^+$~(1.5$\sigma)$,
\item[-]      $D^{+}\to \bar{K}^{*}(700)^0(\rm{POLE})\rho(770)^+$~(3.0$\sigma)$,
\item[-]      $D^{+}\to \bar{K}^{*}(700)^0(\rm{POLE})\rho(1450)^+$~(1.5$\sigma)$,
\item[-]      $D^{+}\to \pi^+\bar{K}_1(1270)^0, \bar{K}_1(1270)^0[D]\to K^-\rho(770)^+$~(3.3$\sigma)$,
\item[-]      $D^{+}\to \pi^+\bar{K}_1(1270)^0, \bar{K}_1(1270)^0[S]\to \bar{K}^*\pi$ ~(\textless1$\sigma)$,
\item[-]      $D^{+}\to \pi^+\bar{K}_1(1270)^0, \bar{K}_1(1270)^0[D]\to \bar{K}^*\pi$ ~(\textless1$\sigma)$,
\item[-]      $D^{+}\to \pi^0\bar{K}_1(1270)^+, \bar{K}_1(1270)^+[S]\to \bar{K}^{*}(892)^0\pi^+$ ~(\textless1$\sigma)$,
\item[-]      $D^{+}\to \pi^0\bar{K}_1(1270)^+, \bar{K}_1(1270)^+[D]\to \bar{K}^{*}(892)^0\pi^+$ ~(3.1$\sigma)$,
\item[-]      $D^{+}\to \pi^+\bar{K}_1(1400)^0, \bar{K}_1(1400)^0[S]\to K^-\rho(770)^+$ ~(2.3$\sigma)$,
\item[-]      $D^{+}\to \pi^+\bar{K}_1(1400)^0, \bar{K}_1(1400)^0[D]\to K^-\rho(770)^+$ ~(1.6$\sigma)$,      
\item[-]      $D^{+}\to \pi^0\bar{K}_1(1400)^+, \bar{K}_1(1400)^+[S]\to \bar{K}^{*}(892)^0\pi^+$ ~(3.9$\sigma)$,
\item[-]      $D^{+}\to \pi^0\bar{K}_1(1400)^+, \bar{K}_1(1400)^+[D]\to \bar{K}^{*}(892)^0\pi^+$ ~(4.5$\sigma)$,
\item[-]      $D^{+}\to \bar{K}^*(1410)^0\pi^+, \bar{K}^*(1410)^0\to \bar{K}^*\pi$ ~(1.9$\sigma)$,
\item[-]      $D^{+}\to \bar{K}^*(1410)^0\pi^+, \bar{K}^*(1410)^0\to K^-\rho(770)^+$ ~(1.3$\sigma)$,
\item[-]      $D^{+}\to \bar{K}^*(1410)^0\pi^+, \bar{K}^*(1410)^0\to \bar{K}^{*}(892)^0\pi^+$ ~(\textless1$\sigma)$,
\item[-] $D^{+}\to \pi^+\bar{K}(1460)^0, \bar{K}(1460)^0\to K^-\rho(770)^+$~($4.7\sigma$),
\item[-] $D^{+}\to \pi^+\bar{K}_1(1650)^0, \bar{K}_1(1650)^0[S,D]\to K^{*}(892)\pi$~($3.0\sigma$),
\item[-] $D^{+}\to \bar{K}^*(1680)^0\pi^+, \bar{K}^*(1680)^0\to K^-\rho(770)^+$~($4.5\sigma$),
\item[-] $D^{+}\to \bar{K}^*(1680)^+\pi^0,  \bar{K}^*(1680)^+\to \bar{K}^{*}(892)^0\pi^+$~($2.3\sigma$).
\item \textbf{Three-body amplitudes}
\item[-] $D^{+}[S]\to \bar{K}^{*}(892)^0(\pi^+\pi^0)_V$~($2.4\sigma$),
\item[-] $D^{+}[P]\to \bar{K}^{*}(892)^0(\pi^+\pi^0)_V$~($2.1\sigma$),
\item[-] $D^{+}[D]\to \bar{K}^{*}(892)^0(\pi^+\pi^0)_V$~($1.6\sigma$),
\item[-] $D^{+}[P]\to \rho(770)^+(K^-\pi^+)_V$~($4.3\sigma$),
\item[-] $D^{+}[D]\to \rho(770)^+(K^-\pi^+)_V$~($2.3\sigma$),
\item[-] $D^{+}[S]\to \rho(1450)^+(K^-\pi^+)_V$~($2.6\sigma$),
\item[-] $D^{+}[P]\to \rho(1450)^+(K^-\pi^+)_V$~($2.4\sigma$),
\item[-] $D^{+}[D]\to \rho(1450)^+(K^-\pi^+)_V$~($2.0\sigma$),
\item[-] $D^{+}\to \rho(770)^+(K^-\pi^+)_{S-\rm wave}$~($2.0\sigma$),
\item[-] $D^{+}\to \bar{K}^{*}(700)^0(\rm{RBW})$$(\pi^+\pi^0)_V$~($3.0\sigma$),
\item[-] $D^{+}\to \bar{K}^{*}(700)^0(\rm{RBW})$$(\pi^+\pi^0)_S$~($1.2\sigma$),
\item[-] $D^{+}\to \bar{K}^{*}(700)^0(\rm{POLE})$$(\pi^+\pi^0)_V$~($2.7\sigma$),
\item[-] $D^{+}\to \bar{K}^{*}(700)^0(\rm{POLE})$$(\pi^+\pi^0)_S$~($0.9\sigma$).

\item \textbf{Four-body non-resonance amplitudes}
\item[-] $D^{+}\to (K^-((\pi^+\pi^0)_{S-\rm wave})_A\pi^+)$~($<1\sigma$),
\item[-] $D^{+}\to (K^-((\pi^+\pi^0)_{S-\rm wave})_P\pi^+)$~($1.6\sigma$),
\item[-] $D^{+}\to (K^-((\pi^+\pi^0)_{S-\rm wave})_V\pi^+)$~($<1\sigma$),
\item[-] $D^{+}\to \pi^0((K^-\pi^+)_{S-\rm wave}\pi^+)_A$~($<1\sigma$),
\item[-] $D^{+}\to \pi^0((K^-\pi^+)_{S-\rm wave}\pi^+)_P$~($<1\sigma$),
\item[-] $D^{+}\to \pi^+((K^-\pi^+)_{S-\rm wave}\pi^+)_V$~($2.3\sigma$),
\item[-] $D^{+}\to \pi^+((K^-\pi^+)_{V}\pi^0)_A$~($<1\sigma$),
\item[-] $D^{+}\to \pi^+((K^-\pi^+)_{V}\pi^0)_P$~($<1\sigma$),
\item[-] $D^{+}\to \pi^+((K^-\pi^+)_{V}\pi^0)_V$~($<1\sigma$),
\item[-] $D^{+}\to (K^-\pi^+\pi^+\pi^0)_{\rm{non-resonance}}$~($3.2\sigma$).

 \end{itemize}
 
 \renewcommand\thesection{\Alph{section}}
\section{The interference between processes}
\label{app:interference}
The interference between processes calculated by Equation~(\ref{interferenceFF-Definition}).

\begin{itemize}
\item[I]  $D^{+}[S]\to \bar{K}^{*}(892)^0\rho(770)^+$,
\item[II] $D^{+}[P]\to \bar{K}^{*}(892)^0\rho(770)^+$,
\item[III] $D^{+}\to \bar{K}_1(1270)^0\pi^+[S]$,
\item[IV] $D^{+}\to \bar{K}_1(1400)^0\pi^+[S]$,
\item[V]  $D^{+}\to \bar{K}_1(1400)^0\pi^+[D]$,
\item[VI] $D^{+}\to \bar{K}(1460)^0\pi^+$,
\item[VII] $D^{+}\to \bar{K}^{*}(1680)^{0}\pi^+$,
\item[VIII] $D^{+}\to (K^-\pi^+)_{S-\rm wave}\rho(770)^+$,
\item[IX] $D^{+}\to \bar{K}(1460)^0\pi^+$, $\bar{K}(1460)^0\to K^-(\pi^+\pi^0)_V$,
\item[X] $D^{+}\to \bar{K}(1460)^0\pi^+$, $\bar{K}(1460)^0\to (K^-\pi)_{V}\pi$,
\item[XI] $D^{+}\to (K^-\rho(770)^+)_A\pi^+$,
\item[XII] $D^{+}\to (\bar{K}^{*}(892)\pi)_A\pi^+$,
\item[XIII] $D^{+}\to (\bar{K}^{*}(892)^0\pi^+)_A\pi^0$,
\item[XIV] $D^{+}\to (K^-(\pi^+\pi^-)_V)_P\pi^+$,
\item[XV] $D^{+}[S]\to (K^-\pi^+)_V\rho(770)^+$.	

 \end{itemize}
 
\begin{table}[htbp]\footnotesize
	\centering
\setlength{\tabcolsep}{1.5mm}{
	\begin{tabular}{c| c c c c c c c c c c c c c c}
	\hline
	\hline
		 &II &III &IV &V &VI &VII &VIII &IX &X &XI &XII &XIII &XIV &XV\\
	\hline$-$
	    I   &$-$0.02 &4.72 &$-$6.74 &3.76 &$-$5.36 &3.85 &$-$4.23 &2.19 &4.54 &$-$2.06 &$-$8.12 &$-$4.19 &8.42 &$-$0.76\\
	    
	    II &      &0.00 &0.00 &0.00 &0.00 &$-$0.87 &0.87 &0.00 &0.01 &0.01 &0.00 &0.00 &0.00 &0.00\\
		
	    III  & & &$-$2.89 &2.84 &$-$0.13 &0.18 &2.89 &$-$3.19 &0.25 &$-$0.40 &$-$0.49 &0.63 &0.26 &$-$0.65\\
	  
             IV  & & &  &$-$1.44 &$-$0.96 &2.40 &$-$2.18 &2.10 &1.30 &3.43 &8.61 &$-$4.11 &$-$2.61 &0.68\\
		
	    V  & & & & &0.07 &$-$0.07 &0.56 &$-$0.58 &0.03 &$-$0.06 &0.35 &$-$0.26 &0.01 &$-$0.24\\
	    
      	    VI  & & & & &  &0.00 &0.03 &2.21 &$-$4.46 &2.63 &$-$0.36 &1.31 &$-$0.79 &$-$0.60\\
      
	    VII  & & & & & & &$-$0.01 &0.02 &0.2 &0.01 &0.00 &0.00 &0.00 &0.00\\
	  
	    VIII  & & & & & & & &3.26 &$-$10.26 &6.52 &$-$1.63 &3.04 &6.66 &$-$6.56\\
	 
      	    IX  & & & & & & & & &$-$5.84 &6.37 &$-$0.14 &0.29 &3.02 &$-$2.45\\
       	    X  & & & & & & & & & &$-$0.74 &0.87 &1.11 &$-$3.74 &3.09\\
             XI  & & & & & & & & & & &$-$0.44 &0.25 &0.45 &0.03\\
             XII & & & & & & & & & & & &1.14 &$-$1.02 &0.30\\
             XIII & & & & & & & & & & & &  &0.67 &$-$0.33\\
             XIV  & & & & & & & & & & & & &  &0.15\\
    \hline
	\hline
    \end{tabular}}
 	\caption{Interference of each amplitude, in unit of \% of total amplitude. }	   
		\label{table:inter}
\end{table}

M.~Ablikim$^{1}$, M.~N.~Achasov$^{4,c}$, P.~Adlarson$^{76}$, O.~Afedulidis$^{3}$, X.~C.~Ai$^{81}$, R.~Aliberti$^{35}$, A.~Amoroso$^{75A,75C}$, Y.~Bai$^{57}$, O.~Bakina$^{36}$, I.~Balossino$^{29A}$
% ...
\\
\vspace{0.2cm}
(BESIII Collaboration)\\
\vspace{0.2cm} {\it
$^{1}$ Institute of High Energy Physics, Beijing 100049, People's Republic of China\\
% ...
$^{3}$ Bochum  Ruhr-University, D-44780 Bochum, Germany\\
$^{4}$ Budker Institute of Nuclear Physics SB RAS (BINP), Novosibirsk 630090, Russia\\
% ...
$^{29}$ INFN Sezione di Ferrara, (A)INFN Sezione di Ferrara, I-44122, Ferrara, Italy; (B)University of Ferrara,  I-44122, Ferrara, Italy\\
% ...
$^{35}$ Johannes Gutenberg University of Mainz, Johann-Joachim-Becher-Weg 45, D-55099 Mainz, Germany\\
$^{36}$ Joint Institute for Nuclear Research, 141980 Dubna, Moscow region, Russia\\
% ...
$^{57}$ Southeast University, Nanjing 211100, People's Republic of China\\
% ...
$^{75}$ University of Turin and INFN, (A)University of Turin, I-10125, Turin, Italy; (B)University of Eastern Piedmont, I-15121, Alessandria, Italy; (C)INFN, I-10125, Turin, Italy\\
$^{76}$ Uppsala University, Box 516, SE-75120 Uppsala, Sweden\\
% ...
$^{81}$ Zhengzhou University, Zhengzhou 450001, People's Republic of China\\

\vspace{0.2cm}
% ...
$^{c}$ Also at the Novosibirsk State University, Novosibirsk, 630090, Russia\\
% ...

}

\end{document}